\documentclass{svjour3}

\usepackage{graphicx}% Include figure files

\begin{document}

\title{Continuous image distortion by astrophysical thick lenses}

\author{Thomas P. Kling         \and
        Louis Bianchini
}

\institute{Thomas P. Kling \and Louis Bianchini \at
              Department of Physics, Bridgewater State University, Bridgewater, MA 02325 \\
              Tel.: +1-508-531-2895\\
              Fax: +1-508-531-1785\\
              \email{tkling@bridgew.edu}  \\
}

\date{Received: date / Accepted: date}

\date{\today}

\maketitle

\begin{abstract}

\noindent Image distortion due to weak gravitational lensing is examined using a non-perturbative method of integrating the geodesic deviation and optical scalar equations along the null geodesics connecting the observer to a distant source.  The method we develop continuously changes the shape of the pencil of rays from the source to the observer with no reference to lens planes in astrophysically relevant scenarios.  We compare the projected area and the ratio of semi-major to semi-minor axes of the observed elliptical image shape for circular sources from the continuous, thick-lens method with the commonly assumed thin-lens approximation.  We find that for truncated singular isothermal sphere and NFW models of realistic galaxy clusters, the commonly used thin-lens approximation is accurate to better than 1 part in $10^4$ in predicting the image area and axes ratios.  For asymmetric thick lenses consisting of two massive clusters separated along the line of sight in redshift up to $\Delta z = 0.2$, we find that modeling the image distortion as two clusters in a single lens plane does not produce relative errors in image area or axes ratio more than $0.5\%$.

\end{abstract}


%________________________________________________________

\section{Introduction} \label{intro:sec}

The analysis of gravitational lensing systems typically proceeds by
invoking a thin-lens approximation, where the three-dimensional matter
distribution responsible for distorting bundles of light rays is
projected into a two-dimensional lens plane.  Sources are assumed to
be located in a source plane, and the mapping of image locations (in
the lens plane) to source locations (in the source plane) is
codified into a lens equation at the heart of nearly all
observational studies.  The Jacobian of the lens equation map
provides the basis for the thin-lens approach to the magnification
and shearing of images or weak gravitational lensing.

Beginning with Frittelli and Newman \cite{fn} and Perlick
\cite{perlick}, a number of authors
developed the theory of gravitational lensing free from lens and
source planes by emphasizing that the null geodesic equations
provide the exact versions of the ``time of flight'' and lens
equations usually invoked in the thin lens approximation.  Around the same time, Pyne and Birkinshaw \cite{pyne} approached the problem of light ray paths and image magnification by perturbatively solving the null geodesic equations.

The exact lens equation approach was first applied to spherically symmetric black hole solutions \cite{fkn:bh}, \cite{ve}.  It has since then been applied in the context of lensing by the rotating black hole at the center of our galaxy \cite{bz}, \cite{bz1}.  Other work examines black hole lensing, including time delays and magnifications, in the context of relativistic images produced by black holes at the centers of galaxies other than our own \cite{virb}. For cosmological applications widely used in observational strong lensing, the accuracy of the thin-lens approximation's predictions of total cluster masses from the appearance of arcs or Einstein rings was determined to be approximately $2\%$ in the absence of a suitable truncation scheme \cite{kf:strong}.

Following this, a series of papers \cite{fkn1}, \cite{fkn2} redefined image
distortion from the exact lens equation. The central idea of these papers was to consider the pencil of rays in an observer's past light cone that connected the observer to an extended source.  The cross-sectional shape of the pencil changes along the past light cone. At the observer, the projection of the cross-section is the observed shape of the source, while at the source, the cross-section of the pencil matches the extended source shape.  The shape of the cross-section of the pencil was determined by integrating the geodesic deviation equations continuously.

These papers reproduced the formalism of the thin-lens approach by introducing a Jacobian of the exact lens mapping.  In the first paper \cite{fkn1}, a set of new definitions for shape parameters and image distortion parameters were proposed that measured the total change of the area or ellipticity (defined as the ratio of semi-major to semi-minor axes) of the cross-section of the pencil of rays.  These image distortion definitions were applied to Schwarzschild black hole geometries.  By considering curvature tensors with support only in a lens plane, the second paper in the series \cite{fkn2} reproduced exactly the thin-lens Jacobian
and discussed the relation between the optical scalars --  the
general relativity ``convergence'' (or divergence), $\rho$, and ``shear,'' $\sigma$ -- and the thin-lens quantities called the same things: $\kappa$, and
$\gamma_1$ and $\gamma_2$.

Frittelli and Oberst \cite{fo} then examined image distortion by thick lenses in simple (non-physical) models where the Ricci curvature was zero and the Weyl curvature was in the form of a square barrier.  This paper integrated the optical scalar equations along straight line paths through the thick barrier and discussed the distortion parameters introduced in \cite{fkn1}.

From the perspective of the astrophysicist analyzing the statistical, weak-lensing signal from real systems, these three papers examining image distortion suffer from three significant short-comings.  First, the shape and image distortion parameters introduced in \cite{fkn1} do not correspond in a natural way to the actual observable quantities.  Second, the curvature models considered in these papers are not appropriate for modeling galaxy clusters.  Finally, the completely general choice of basis vectors (which was necessary for examining image distortion in general, say near a black hole) may prevent the reader from seeing how one would apply the formalism to typical data where the gravitational perturbation is weak.

Like the work begun in \cite{fkn1}, the purpose of the current paper is to re-examine image distortion purely in terms of the physical processes of continuous distortion that occur to a pencil of light rays as the pencil travels from the source to the observer -- with no reference to lens planes.  Unlike the work of Pyne and Birkinshaw, we will solve the underlying geodesic and geodesic deviation equations exactly.   We will specifically assume typical statistical weak lensing conditions, including that the gravitational potential provides a weak perturbation of a background Robertson-Walker metric.  We integrate the optical scalar and geodesic deviation equations in this context and state our results in the language that is customary in the observational lensing community.

The standard references, e.g. \cite{ehlers} and \cite{schneider}, provide a very brief and general discussion of the geodesic deviation equations as the underlying physical description of image distortion.  This paper explicitly formulates in a direct manner the simplest method of integrating the geodesic deviations equations using the Newman-Penrose spin coefficient formalism applied in the context of weak gravitational lensing.

Specifically, we will begin with curvature tensors representing
potentially realistic matter distributions, and derive expressions for the
general relativity convergence, $\rho$, and shear, $\sigma$, of
pencils of light rays. Then we will integrate the effect of the
convergence and shear along the pencil from the source to the
observer -- finding the size and shape of the image an observer sees
of the object.

While this approach is generally well known within the general relativity community, its explicit application to image distortion in the case of weak gravitational lensing is new.  For example, the main equations for image distortion in this method are found in \cite{perlick}, but the boundary conditions used there are difficult to apply to the weak lensing case because the shape of the pencil is not known a-priori at the observer.

Because our curvature tensors, convergence, and shear are allowed
support throughout all of space-time, the approach to image
distortion that we will derive represents lensing by ``thick
lenses.''  The study of thick lenses is becoming increasingly
important as more examples of merging clusters are found and analyzed, for instance in \cite{okabe}.  The physics of merging clusters has important consequences for dark matter models, galaxy formation and stellar history, and x-ray modeling.

Our methods also allow us to begin to examine the accuracy of the thin-lens approximation in statistical weak gravitational lensing for reasonable lensing configurations.  We examine the differences in predicted image area and axes ratio for a range of lens models.  We also perform preliminary investigations of weak gravitational lensing by two clusters, closely separated along the line of sight by a varying amount in redshift space.

%_________________________________________________

\section{Perturbed metric} \label{metric:sec}

We assume that we have an observer and a series of sources co-moving
in a perturbed flat Robertson-Walker (RW) space-time.  Null geodesics
connect the sources to the observer and are received simultaneously
at the observer.  These null geodesics pass through a region of
space-time where the space-time metric is perturbed by a weak
gravitational potential, $\varphi$.  We make the customary
assumption that $\varphi$ depends on proper distances at the time
the light rays pass the lens.  For instance, if the potential is spherically symmetric, $\varphi$ will depend on $r_p = a(\tilde t_l) r$, where $\tilde t_l$ is the time all the null geodesics of the observer's past light cone pass the lens. In this paper, as elsewhere, we assume $a(\tilde t_l)$ can be treated as a constant since the scale factor varies slowly compared to the time light passes across the lens \cite{ehlers}.  The perturbed RW metric is

\begin{equation}  d \tilde{s}^2 = (1+2 \varphi) \, d\tilde{t}^2
- a(\tilde t)^2 (1-2\varphi) \{ dx^2 + dy^2 + dz^2 \} , \label{m1}
\end{equation}

\noindent where $\tilde t$ is the usual co-moving time and $(x,y,z)$ are cartesian coordinates.  Throughout the paper, we will work in units where $G=c=1$.  Using the conformal time, the metric, Eq.~\ref{m1}, is conformal to the static
metric

\begin{equation}  ds^2 = (1+2 \varphi) \, dt^2
- (1-2\varphi)  \{ dx^2 + dy^2 + dz^2 \} . \label{m2} \end{equation}

\noindent This static metric is a perturbation of a flat,
Minkowski-space background metric.

For pressureless matter, the gravitational potential satisfies

\begin{equation} \nabla^2_p \varphi = 4 \pi G \rho_m, \end{equation}

\noindent where $\rho_m$ is a matter density that depends on the proper distance, and the gradient operator is taken with respect to the same proper distance:  $\nabla_p = a(\tilde t_l)^{-1} \nabla$.

While the cosmology is ``hidden'' in Eq.~\ref{m2}, we will make use of the cosmological choice in defining the spatial positions of the lens, source and observer as in \cite{kf:strong}.  We take the background to be a flat RW metric with the Hubble constant $H_0 = 70$~km/s/Mpc, the matter density $\Omega_m = 0.3$ and the cosmological constant density $\Omega_\Lambda = 0.7 = 1- \Omega_m$. In this, the scale factor is  \cite{ryden}

\begin{equation} a(\tilde t) = \left( \frac{\Omega_m}{\Omega_\Lambda}
\right)^{1/3} \left\{ \sinh\left( \frac{3 H_0 \sqrt{\Omega_\Lambda}
\tilde t}{2} \right) \right\}^{2/3}. \label{a1}  \end{equation}

\noindent We assume the lens is at a redshift $z_l$ while any sources are at redshifts $z_s$.  We use the redshift relation, $ 1+\tilde z =
\frac{a(\tilde t_o)}{a(\tilde t_e)}, $ where $\tilde t_e$ is the time a light ray is emitted that is received at an observer at $\tilde t_o$. Setting $a(\tilde t_o)=1$, we can solve for the value of $\tilde t_e$ with Eq.~\ref{a1}.

Then we obtain the radial positions of the source and observer by orienting
our coordinates in such a way that a light ray travels radially in
the background spacetime and assuming that the observer, lens and
source are at least nearly co-linear. Integrating radial
null geodesics of the flat RW metric, Eq.~\ref{m1}, determines the radial
positions of the source and observer by ignoring the perturbation
introduced by the lens.

%___________________________________________________________

\section{Curvature tensors} \label{NP:sec}

In this section, we specify the approximations that apply our formalism to the case of ground-based, statistical weak lensing studies.  In typical studies, one finds an average background redshift of all the sources to compute the projected matter density.  This is because the measured gravitational shear is related to the dimensionless projected matter density $\kappa = \Sigma / \Sigma_{crit}$ where $\Sigma$ is the mass density of the lens projected along the line of sight and

\[ \Sigma_{crit} = \frac{c^2 D_s}{4 \pi G D_l D_{ls}}, \]

\noindent for angular diameter distances to the lens, source, and between the lens and source.  In practice, an average background $\Sigma_{crit}$, or an average background redshift, is derived from averaging color information over all the sources to specify an average $D_s$ and $D_{ls}$ in $\Sigma_{crit}$.

In this paper, we assume that the matter distribution is localized near the origin, and that the observer lies at some position along the $+\hat z$
axis.  Based on the common practice of finding an average background redshift, we take sources to lie at the same redshift in a source plane at some fixed value of $z<0$ along the $-\hat z$ axis with $x^2+y^2 \ll z^2$.  The observer and sources are far from the region where the gravitational perturbation is large.

The Newman-Penrose (NP) spin coefficient formalism \cite{NP} is based on a tetrad of null vectors. Our use of the tetrad vectors is to create complex scalar components of the Ricci and Weyl tensors by contracting the curvature tensors with the tetrad vectors.  The NP tetrad vectors are customarily denoted by

\begin{equation}\lambda^a_i = \left \{ \ell^a, n^a, m^a, \bar m^a \right \}
\label{tetrad} , \end{equation}

\noindent where the first tetrad vector, $\ell^a$, is tangent to the past-directed null geodesic connecting the source and observer.  The vector $m^a$ is a complex spatial vector parallel propagated along $\ell^a$ that can be taken to lie in the cross-section of a bundle of light rays surrounding $\ell^a$.

Two NP components of the Ricci and Weyl tensors drive image distortion.  The Ricci tensor component is

\begin{equation} \Phi_{00} = -\frac{1}{2} R_{ab} \ell^a \ell^b, \label{ric} \end{equation}

\noindent and the Weyl tensor component is

\begin{equation} \Psi_0 = -C_{abcd} \ell^a m^b \ell^c m^d . \label{wyl} \end{equation}

In principle, the exact tetrad vectors that are null in the perturbed metric are used to find $\Phi_{00}$ and $\Psi_0$. However, we wish to work consistently to first order in the gravitational perturbation.  Since $R_{ab}$ and $C_{abcd}$ are first order in the gravitational potential, we must use an approximate tetrad that is null in the flat background metric.  If $(\zeta, \bar \zeta)$ are complex stereographic coordinates that span a sphere's worth of directions, the cartesian coordinate components of the approximate tetrad vectors can be written as

\begin{eqnarray} \ell^a &=& \frac{1}{\sqrt{2}(1+\zeta\bar\zeta)} \left(
-1-\zeta\bar\zeta, \zeta + \bar\zeta, i(\bar \zeta - \zeta),
-1+\zeta\bar\zeta \right), \label{l} \\
n^a &=& \frac{1}{\sqrt{2}(1+\zeta\bar\zeta)} \left(
1+\zeta\bar\zeta, -(\zeta+\bar\zeta), i(\zeta-\bar\zeta),
1-\zeta\bar\zeta \right), \label{n} \\
m^a &=& \frac{1}{\sqrt{2}(1+\zeta\bar\zeta)} \left( 0,
1-\bar\zeta^2, -i (\bar\zeta^2 +1), 2\, \bar\zeta \right). \label{m}
\end{eqnarray}

\noindent For $\zeta = \bar \zeta = 0$, the spatial part of $\ell^a$  points parallel to the $\hat z$ axis towards negative $z$ values.  The approximate tetrad vectors are allowed to vary along the null geodesic from the source to the observer, or $\zeta(s)$ where $s$ is a parameter along the geodesic with $s=0$ at the observer.

 In the context of statistical, wide-field weak gravitational lensing, one studies the distortion of source galaxies that lie behind and ``near'' a large cluster of galaxies that serves as a lens.  With a typical dithering of the stacked images from a 4-m telescope, ground-based studies see approximately 15 arc minutes to each side of the lensing cluster.  This means that $\zeta$ at the observer is small, and, even if it varies along the null geodesic, $\zeta(s)$ remains small from the source to the observer.

For the weakly perturbed metric, Eq.~\ref{m2}, and approximate null
tetrad vectors, Eqs.~\ref{l} and \ref{m}, Kling and Campbell \cite{kc} show that the curvature tensors are related to the gravitational potential by

\begin{equation} \Phi_{00} = \frac{1}{2} \nabla^2 \varphi, \label{ricci} \end{equation}

\noindent and

\begin{equation} \Psi_0 = \frac{1}{2} ( \varphi_{xx} -
\varphi_{yy} - 2i\varphi_{xy} ) , \label{weyl} \end{equation}

\noindent keeping the first order terms of $\varphi$ to be
consistent with the weak-field approximation in the metric.  All derivatives are taken with respect to the cartesian coordinates, and $\varphi_{xx} = \partial ^2 \varphi / \partial x^2$.  The
Ricci tensor, Eq.~\ref{ricci}, is exact for all values of $\zeta$,
but the Weyl tensor assumes $\zeta$ to be small and discards terms
of order $\zeta \varphi$ or smaller.  That the Ricci and Weyl tensor components do not depend $\zeta$ implies that these are dependent only on the position in space $(x(s), y(s), z(s))$ along the light ray path for a static perturbation.

%________________________________________________________

\section{Geodesic deviation equations} \label{gd:sec}

The null geodesic connecting the source and observer is
part of the observer's past light cone.  Within this light cone is a pencil of rays that surrounds the null geodesic.  The pencil of rays collapses to a point at the observer (or the apex of the light cone).  The shape of this pencil at the source corresponds to the true shape of the source.

The rate of change of the shape of the pencil is determined by the
optical scalars: $\rho$ and $\sigma$.  The divergence, $\rho$,
measures the rate of the expansion of the pencil.  The complex
shear, $\sigma$, measures the rate of change of the ``shearing'' of
the pencil, or the tendency of the cross section of the pencil to
become elliptically shaped.

If one integrated the null geodesic equations of the
perturbed metric, Eq.~\ref{m2}, one could find an exact null tetrad $\{ \ell^a, n^a, m^a, \bar m^a \}$.  In the spin coefficient formalism, the
divergence and shear can be computed from the exact null tetrad as

\begin{equation} \rho = \ell_{a;b}m^a\bar m^b
\quad\quad \sigma = \ell_{a;b}m^a m^b. \label{os1} \end{equation}

Alternately, one can compute the optical scalars directly from the
curvature tensors $\Phi_{00}$ and $\Psi_0$.  Using $s$ as the parameter along the null geodesic whose tangent vector is $\ell^a$, $D = \frac{1}{\sqrt{2}} \frac{d}{ds}$ is the directional derivative along $\ell^a$.  Then the Sachs equations specify how the optical scalars evolve along the null geodesic whose tangent vector is $\ell^a$:

\begin{eqnarray} D \rho &=& \rho^2 + \sigma \bar \sigma + \Phi_{00},
\label{sachs:rho} \\ D \sigma &=& 2\rho\sigma + \Psi_0.
\label{sachs:sigma} \end{eqnarray}

\noindent We will treat Eqs.~\ref{sachs:rho} and \ref{sachs:sigma} as ordinary differential equations for $\rho$ and $\sigma$ where $\Phi_{00}$ and $\Psi_0$ are source terms (functions of $s$) evaluated at positions $(x(s), y(s), z(s))$ along the null geodesic path.

Following Penrose and Rindler \cite{PR}, we introduce a real Jacobi vector that is carried along the null geodesic $\ell^a$ and is in the cross-section of the pencil of rays:

\begin{equation} q^a = \xi \bar m^a + \bar \xi m^a , \end{equation}

\noindent where $m^a$ is the background tetrad vector. Note that since $\zeta$ is small

\begin{equation} m^a \approx \frac{1}{\sqrt{2}} \left( 0,
1 , -i , 0 \right) \equiv \frac{1}{\sqrt{2}} \left( e_x^a - i e_y^a
\right)
\end{equation}

\noindent for a pair of constant, real, unit vectors $e_x^a$ and
$e_y^a$ that point in the $+\hat x$ and $+\hat y$ directions,
respectively.  Taking $\xi = (a-ib)/\sqrt{2}$, the Jacobi field is related to the real basis vectors as

\begin{equation} q^a = a e_x^a + b e_y^a  \label{jacobi} \end{equation}

\noindent for small $\zeta$.
Writing $Z = (\xi, \bar \xi)$, the geodesic deviation
equation is equivalent to

\begin{equation} D Z = -P Z, \label{geodev}\end{equation}

\noindent for

\begin{equation} P = \left( \begin{array}{cc} \rho & \sigma \\ \bar \sigma & \rho \end{array} \right). \end{equation}

Because our bundle of light rays converges at the observer, the divergence, $\rho$, is real. However, the shear, $\sigma$ is not in general.  Writing $\sigma = \sigma_r + i \sigma_i$, the real and imaginary parts of the geodesic deviation equation separate into two real equations:

\begin{eqnarray} D a &=& -(\rho + \sigma_r) a + \sigma_i b, \nonumber
\\ D b &=& -(\rho - \sigma_r) b + \sigma_i a. \label{geodev1}
\end{eqnarray}

The procedure we will use for integrating the geodesic deviation vectors will be to first find the optical scalars from Eqs.~\ref{sachs:rho} and \ref{sachs:sigma} which are governed by the curvature tensors.  Then we will integrate the two real components of the geodesic deviation vector in Eqs.~\ref{geodev1}.  The integration of the optical scalars will proceed backwards in time from the observer to the source along the past-directed null geodesic from the observer to the source.  However, because we are ultimately interested in understanding the observed shape of the source at the observer, we will integrate the geodesic deviation vectors from the source towards the observer (forwards in time).  Since we are interested in the observational case of the observed shape of intrinsically (on-average) circular sources, our approach in integrating the geodesic deviation vectors differs from the presentation in typical references in the general relativity community, for example in \cite{perlick}.

%________________________________________________________

\section{Flat space} \label{flatspace:sec}

We first consider flat space ($\varphi=0$) to understand the boundary conditions we should impose on the integration of $\rho$, $\sigma$ and the geodesic deviation vectors. This will also motivate our definition of the area and ratio of semi-major to semi-minor axes of elliptical images of the source at the observer.

Here, the curvature tensors are zero, and the spatial part of the null geodesics are simply straight lines.  Because there is no curvature, the ``lens'' has axial symmetry about the $\hat z$ axis and we can take a null geodesic whose spatial part is in the $\hat x$-$\hat z$ plane as generic.

The solutions to the Sachs equations consistent with the pencil forming part of the past light cone of the observer are

\begin{equation} \rho = -\frac{1}{\sqrt 2 s}, \quad\quad \sigma = 0 \label{flatscalars}. \end{equation}

\noindent The meaning of the divergence of $\rho$ at $s = 0$ is that the rays forming the pencil have converged at the observer.

The geodesic deviation equation, Eq.~\ref{geodev1} becomes

\begin{equation} \dot a = \frac{a}{s} \quad\quad
 \dot b = \frac{b}{s}, \label{flatgeodev}
\end{equation}

\noindent where $\cdot = d/ds$.  We set the boundary condition at the source ($s=s_f$) to be $a(s_f) = \alpha$, $b(s_f) = \beta$ with $\alpha^2+\beta^2 = 1$.  The solution to the geodesic deviation equation is then

\begin{equation} a = \frac{\alpha s}{s_f} \quad\quad b = \frac{\beta s}{s_f}.  \label{flatab} \end{equation}

\noindent We see that the geodesic deviation vector in flat space vanishes at the observer, as it should, because the pencil of rays has converged there.

To model a circular source, we take a pair of geodesic deviation vectors, $q^a$ and $\tilde q^a$, or $(a, b)$ and $(\tilde a, \tilde b)$, specified by $(\alpha = 1, \beta = 0)$ and $(\tilde \alpha = 0, \tilde \beta = 1)$. For a null geodesic whose spatial projection is in the $\hat x$-$\hat z$ plane, $q^a$ points in the $+\hat x$ direction and $\tilde q^a$ points in the $+\hat y$ direction along the entire geodesic, and this pair of geodesic deviation vectors uniquely determines the circular cross-section of the bundle along the null geodesic from the observer to the source.

We define the projected ratio of semi-major to semi-minor axes, or axes ratio, to be

\begin{equation} R_p = \lim_{s \rightarrow 0} \left( \frac{\tilde b} {a} \right) \label{ratio:def}, \end{equation}

\noindent  and we define the projected area of the image ellipse to be

\begin{equation} A_p = \lim_{s \rightarrow 0} \left\{a \tilde b \left( \frac{s_f^2}{s^2} \right) \right\} \label{area:def}, \end{equation}

\noindent In flat space, both the projected area and projected ratio defined this way are $1$, which means that the image of the circular source that an observer sees is a circle of scaled unit area.  These definitions for the projected area and ratio hold for any axially symmetric lens and will allow for comparison with similarly defined quantities in the traditional thin-lens approximation.

%___________________________________________________________

\section{Optical scalars with non-zero curvature} \label{opticals:sec}

To consider integrating the optical scalar equations with non-zero curvature, we introduce the variable $u = 1/\rho$.  The optical scalar equations read

\begin{eqnarray} \dot u = -\sqrt{2} ( 1 + u^2(\sigma\bar\sigma + \Phi_{00})), \label{udot} \\ \dot \sigma = 2\sqrt{2} \left(\frac{\sigma }{u}\right) + \sqrt{2} \Psi_0. \label{sigmaudot} \end{eqnarray}

\noindent We assume that the observer is far from any region of significant curvature, so that the initial past light cone is a light cone in flat space. The boundary conditions consistent with the pencil of rays forming part of the flat-space, past light cone are $u(s = 0) = 0$, $\sigma (s=0) = 0$, and

\[ \lim_{s \rightarrow 0} \frac{\sigma}{u} = 0. \]

\noindent This last condition insures that if the curvature tensors are zero, one obtains the flat space $\sigma = 0$ solution.  These conditions allow for a well defined numerical integration of $u$ and $\sigma$ near $s = 0$.  We note from the minus sign in Eq.~\ref{udot} that $u$ will be negative near the observer -- so that $\rho \approx -1/(\sqrt{2}{s})$ near the observer.

At some point after $s = 0$, it is sometimes convenient to switch back into the $(\rho, \sigma)$ pairing:

\begin{eqnarray} \dot \rho = \sqrt{2} (\rho^2 + \sigma\bar\sigma + \Phi_{00}), \label{rhodot} \\ \dot \sigma = 2\sqrt{2} \rho\sigma  + \sqrt{2} \Psi_0. \label{sigmadot} \end{eqnarray}

\noindent This is especially true because $\rho$ can become zero or $u\rightarrow -\infty$, forcing the switch for numerically integrating the equations.

A general property of pencils of light rays that encounter non-zero curvature at some point along the pencil is that conjugate points will form if the geodesic can be extended far enough \cite{wald}.  Conjugate points are places where a Jacobi vector vanishes at a value of $s \ne 0$ -- so that the Jacobi vector vanishes in two places (at the observer and somewhere else).

If the weak energy condition $\Phi_{00}>0$ holds, from Eq.~\ref{rhodot}, we see that $\rho$ is always increasing.  In fact, $\rho$ can become zero, meaning that the beam has stopped expanding, and then $\rho$ will eventually run away to positive infinity with $\sigma$ running away to negative infinity as the beam contracts.

\begin{figure}[h]
\begin{center}$
\begin{array}{cc}
\includegraphics[width=2.5in]{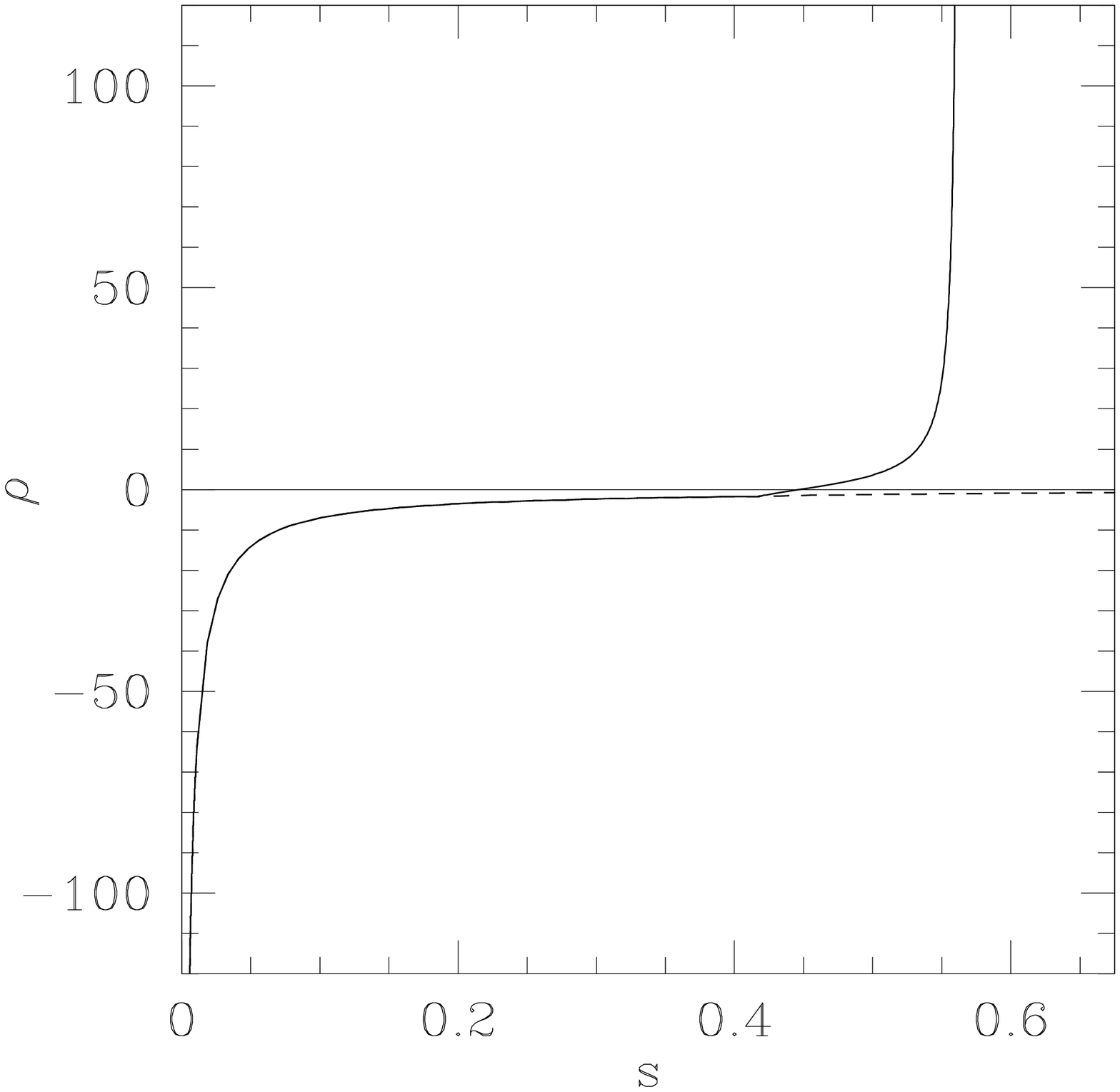} &
\includegraphics[width=2.5in]{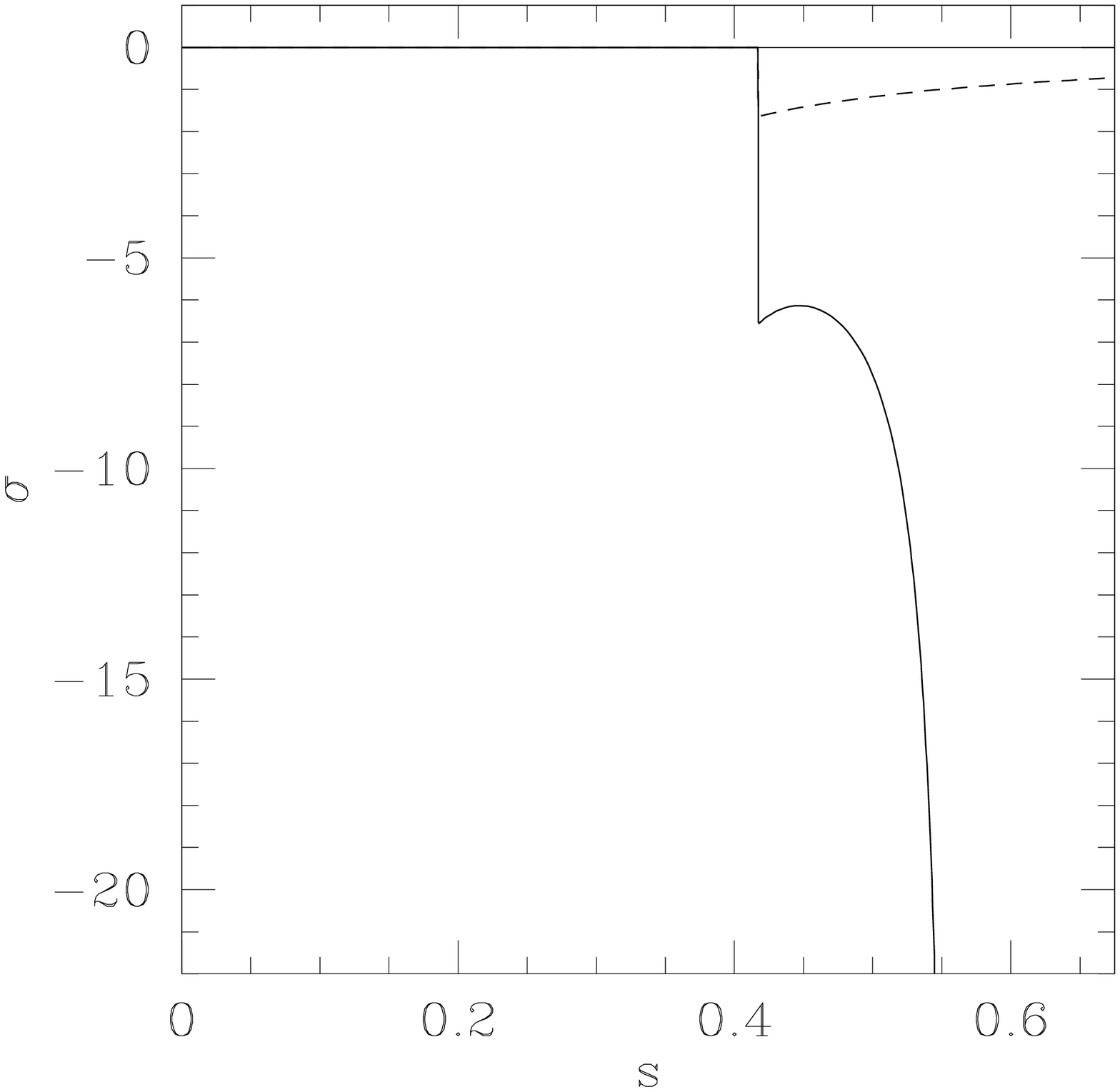}
\end{array}$
\end{center}
\caption{The optical scalars, $\rho$ and $\sigma$, along null geodesics in a space-time perturbed by a point mass potential with a total mass of $2\times10^{15}$ solar masses for $\theta= 60$ arc sec (solid line) and $\theta = 120$ arc sec (dashed line).  The point mass is placed at a redshift of $0.45$.  A conjugate point is reached along the $60$ arc sec null geodesic where the optical scalars diverge. \label{rhosig:pm} }
\end{figure}

\begin{figure}[h]
\begin{center}$
\begin{array}{cc}
\includegraphics[width=2.5in]{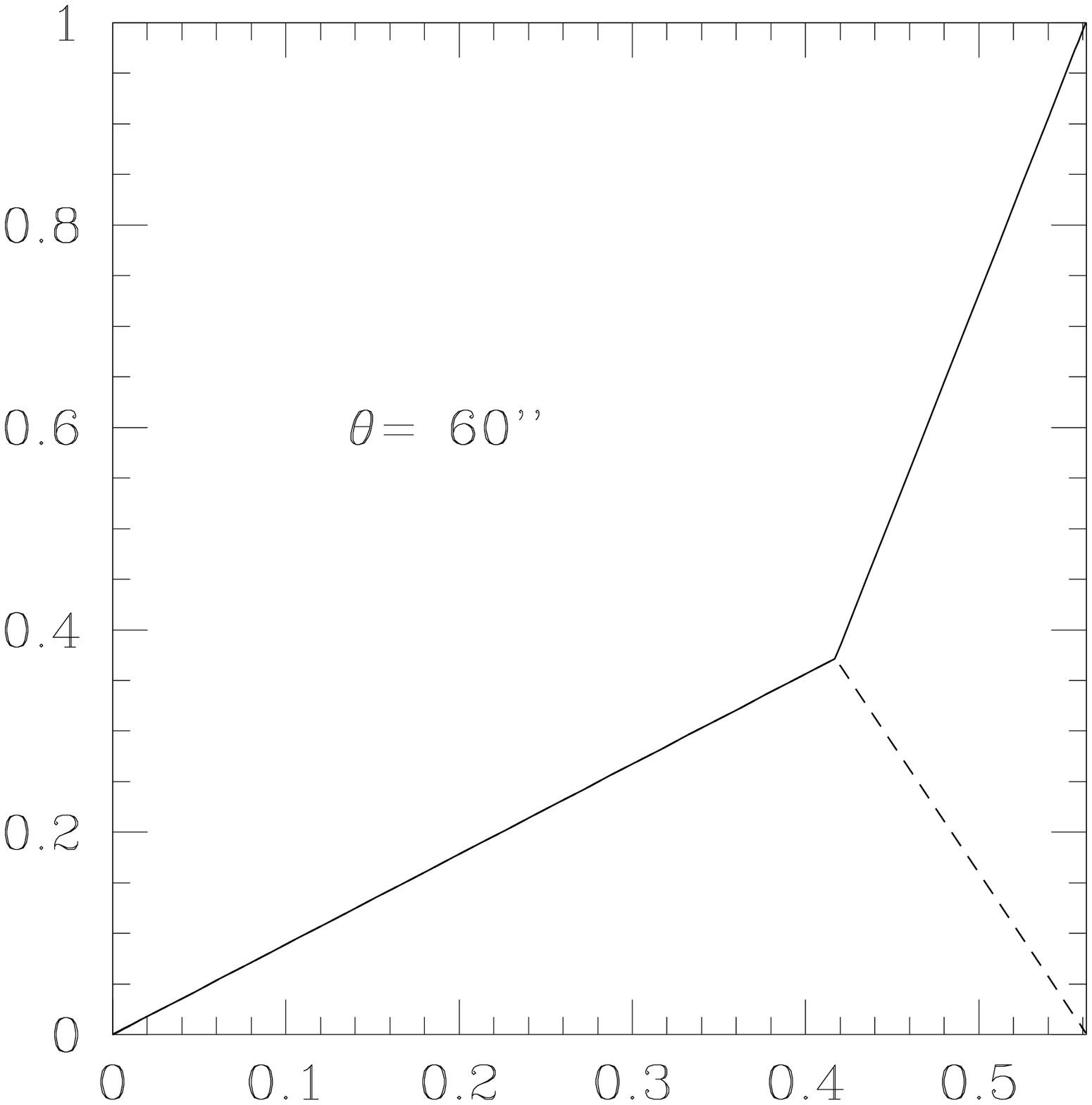} &
\includegraphics[width=2.5in]{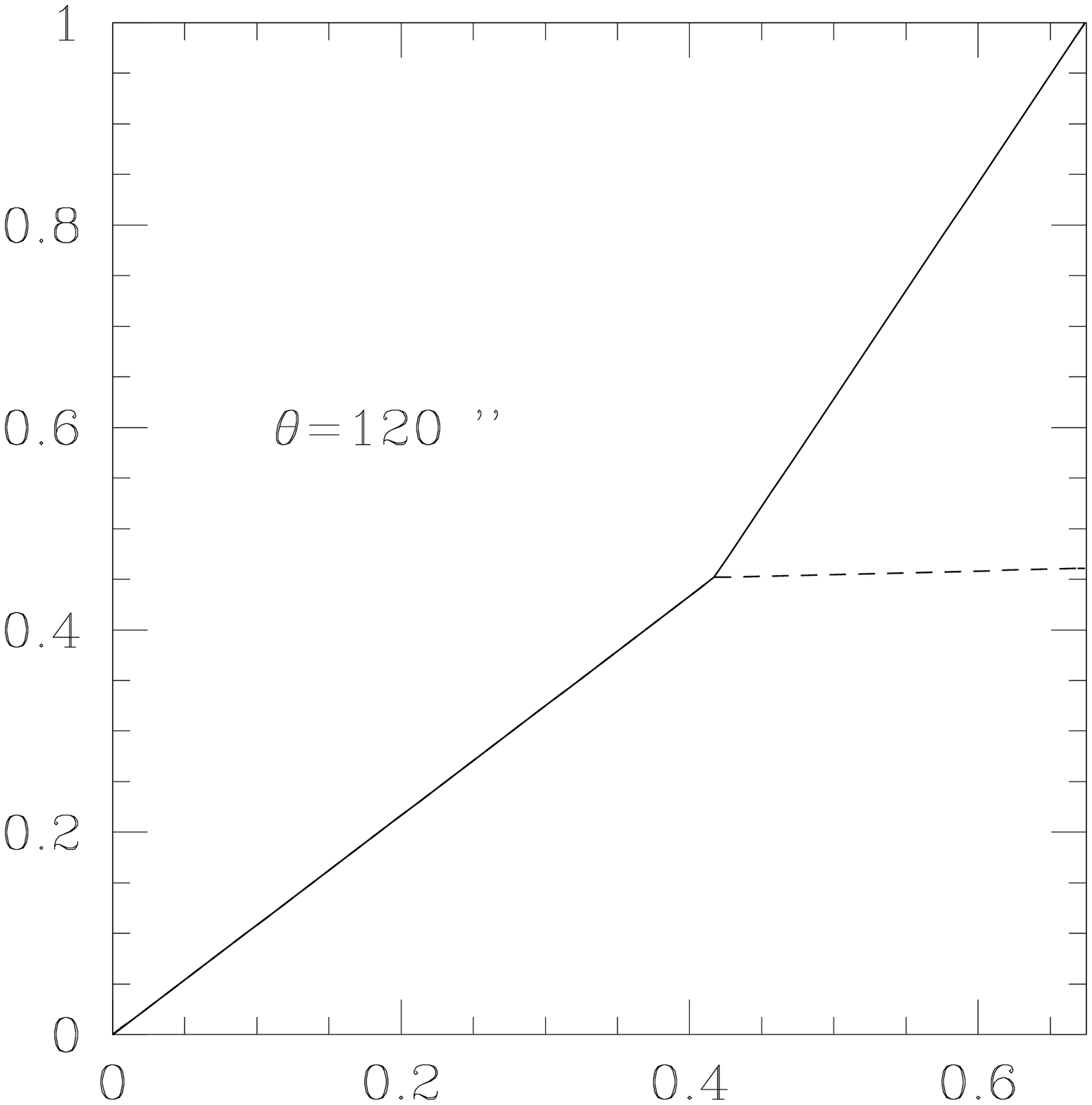}
\end{array}$
\end{center}
\caption{Geodesic deviation vector components $a$ (solid) and $b$ (dashed) along null geodesics in a space-time perturbed by a point mass potential with a total mass of $2\times10^{15}$ solar masses for $\theta= 60$ and $120$ arc sec.  The point mass is placed at a redshift of $0.45$. The vanishing of the $b$ component along the $\theta=60$ arc sec null geodesic indicates that a conjugate point has been reached.  \label{ab:fig} }
\end{figure}

Figure~\ref{rhosig:pm} shows the behavior in $\rho$ and $\sigma$ in the case of a point mass curvature for two null geodesics: one making an angle of $\theta = 60$~arc sec from the optical axis and a second making an angle of $\theta = 120$~arc seconds. In the case of the $60$ arc sec ray, the pencil has reached a conjugate point before $s=0.55$.  Figure~\ref{ab:fig} shows the integration of two geodesic deviation vector components, $(a, b)$, backwards in time subject to the boundary condition that

\[ \lim_{s\rightarrow 0} \frac{a}{s} = \lim_{s\rightarrow 0} \frac{b}{s} = 1 \]

\noindent for $s=0$ at the observer.  The components are scaled in each case such that $a=1$ at a final $s$ value.  In the case of $\theta=60$ arc sec, when the conjugate point is reached, the $b$ component of the geodesic deviation vector has vanished.

The presence of conjugate points presents significant numerical problems to integrating the optical scalar equations.  (The main difficulty is an ambiguity in how to reset the boundary conditions for $(\rho, \sigma)$ on the other side of the conjugate point.)  Conjugate points are common in lensing when one considers rays that pass close to the center of a galactic cluster, for instance with Einstein rings.  Fortunately for the analysis of weak lensing (wide-angles) done on RW cosmological backgrounds with reasonable matter distributions representing galaxy clusters, conjugate points tend not to form until very, very far beyond the position of an observable lensing source.  (The shape of a galaxy with redshift more than 1 is very difficult to measure accurately because one only detects light from the central region of the galaxy.)

%___________________________________________________________

\section{Null geodesic equations} \label{path:sec}

The optical scalar and geodesic deviation equations, Eqs.~\ref{sachs:rho}, \ref{sachs:sigma}, and \ref{geodev1}, are to be integrated along the path the light ray takes through the space-time -- along the null geodesic of the metric, Eq.~\ref{m2}.  The simplest way to obtain the null geodesic equations is to write out the Euler-Lagrange equations of the Lagrangian

\begin{equation} \mathcal{L} = \frac{1}{2} g_{ab} \dot x^a \dot x^b = \frac{1}{2} \left((1+2 \varphi) \, \dot t^2
- (1-2\varphi)  \{ \dot x^2 +\dot y^2 + \dot z^2 \}  \right) = 0. \label{lagrangian} \end{equation}

\noindent The Lagrangian is zero because the geodesics we seek are null.  Working to first order in $\varphi$ and using ${\mathcal{L}}=0$, the spatial equations of motion are

\begin{equation} \ddot x^i = 2 \dot x^i\, (1+2\varphi)(\varphi_{,k} \dot x^k) - 2 \delta^{ij} \varphi_{,j}, \label{eqns:motion} \end{equation}

\noindent where $\varphi_{,i} \equiv \partial \varphi / \partial x^i$.

For an individual past-directed null geodesics, the boundary conditions at the observer, $s = 0$, are $x(0)=y(0)=0$, $z(0) = z_0>0$.  The initial $z_0$ is determined as the radial coordinate distance from the lens (at the origin) when the redshift of the lens is given using Eq.~\ref{a1}.  One can set the initial values of $(\dot x,\dot y, \dot z)$ to be $(\sin\theta\cos\phi, \sin\theta\sin\phi, - \cos\theta)$ for angles on the observers sky $(\theta, \phi)$ where $\theta = 0$ corresponds to looking directly at the lens along the $\hat z$ axis and $\phi$ is a polar coordinate around the $\hat z$ axis.  (The angle of observation for the center of the image between the $\hat z$ axis and the light-ray is $\theta$; the polar angle $\phi$ measures in the ``sky plane'' an angle from the $\hat x$ axis.)

%___________________________________________________________

\section{Thin-lens image distortion} \label{TL:sec}

Following Falco, Schnieder and Ehlers \cite{ehlers}, the thin-lens approximation involves projecting the mass density into a two-dimensional lens plane perpendicular to the optical axis and considering a lens mapping from image locations to a source position.  The lens mapping depends on the bending angle, which is the gradient of the projected gravitational potential

\[ \psi(\vec x) = \frac{1}{\pi} \int_{R^2} \, d^2x' \, \ln{| \vec x - \vec x'|} \kappa(\vec x') \]

\noindent where the integral is taken over the lens plane.

The Jacobian of the lens mapping

\begin{equation} {\mathcal{A}} =  \left( \begin{array}{cc} 1 - \kappa - \gamma_1 & -\gamma_2 \\ -\gamma_2 & 1 - \kappa + \gamma_1 \end{array} \right), \end{equation}

\noindent for the ``shears'' $\gamma_1 = (\psi_{xx} - \psi_{yy})/2$ and $\gamma_2 = \psi_{xy}$, controls weak lensing image distortion.  $\mathcal{A}^{-1}$ can be used to map small vectors from the source plane to the lens (or image) plane. The eigenvectors of $\mathcal{A}^{-1}$ indicate the directions of principle stretching and contraction due to lensing, and the eigenvalues give the magnitudes of the lengths of the semi-axes.  If $\gamma = \sqrt{\gamma_1^2 + \gamma_2^2}$, then a circular source with unit area is mapped into an ellipse where the ratio of the semi-axes is

\begin{equation} R_{tl} = \left| \frac{1-\kappa + \gamma} {1-\kappa - \gamma} \right| \label{tlratio}, \end{equation}

\noindent and the area of the ellipse is

\begin{equation} A_{tl} = \left|\frac{1}{(1-\kappa)^2 - \gamma^2} \right| \label{tlarea}. \end{equation}

%___________________________________________________________

\section{Axially symmetric examples}\label{axial:sec}

In this section, we compare the thick-lens and thin-lens predictions for image area and axes ratio for three axially symmetry matter distributions.  We require each matter distribution to have a finite total mass when integrated over all space.  The matter distributions we examine are a point mass model, a singular isothermal sphere (SIS) that undergoes hard truncation (matter density set to zero for $r>r_c$), and a smoothly truncated version of the model of Navarro, Frenk, and White (NFW) \cite{baltz}.  For shorthand, we will refer to our truncated SIS model as hSIS (reminding the reader of the hard truncation) and the smoothly truncated NFW model as tNFW.

In the next subsection, we briefly outline the three-dimensional potentials and the Ricci and Weyl tensor components used in this paper.  The formulas for the thin-lens projected matter densities, $\kappa$, and shears, $\gamma$, are found in \cite{ehlers}, \cite{kf:strong}, and \cite{baltz} for the point mass, hard truncation SIS, and smoothly truncated NFW.  In following subsections, we discuss the integration of the optical scalar equations and the appearance of conjugate points, and we compare the thin and thick lens image areas and ratios for these axially symmetric matter distributions.

\subsection{Thick lens models}

For the point mass model, the gravitational potential is

\begin{equation} \varphi = -\frac{M}{a_l \sqrt{x^2+y^2+z^2}}, \end{equation}

\noindent where including $a_l = 1/(1+z_l)$ implements the customary approximation that the potential depend on the proper distance at the time the light passes the lens \cite{ehlers}.  The Ricci tensor, $\Phi_{00}$, is zero and the Weyl tensor is

\begin{equation} \Psi_0 = -\frac{3M (x-iy)^2}{2a_l (x^2+y^2+z^2)^{5/2}}. \end{equation}

We take the mass of the point mass model to be $2.0 \times 10^{15}$ solar masses ($M_\odot$) and place the lens at $z_l = 0.45$ and sources at $z_s = 0.8$.  The cluster mass and redshift values chosen here closely correspond to the x-ray luminous cluster RXJ1347-1145, and the source redshift corresponds to the average source redshift in one ground-based observation of that cluster \cite{rxj}.

As discussed in Kling and Frittelli \cite{kf:strong}, singular isothermal sphere models are unphysical without truncation because the total mass of the model $(r<\infty)$ is infinite. The simplest way to truncate is hard truncation: $\rho=0$ for a proper radius greater than some cut-off radius, $r_c$, which is taken as a constant proper distance.  The gravitational potential is given by

\begin{equation} \varphi = \left\{ \begin{array}{ccr} 2
\sigma_v^2 \ln x_p - 2 \sigma_v^2 & \quad\quad & x_p<1 \\
-\frac{2\sigma_v^2}{x_p} & \quad\quad & x_p>1 \end{array} \right. ,
\label{potential_sis} \end{equation}

\noindent where $x_p = a_l r/r_c$ is a natural dimensionless radial
parameter and $\sigma_v$ is the velocity dispersion -- the usual parameter in one-parameter SIS models.  At $r_c$, the model has been continuously tied to a point mass model with the same total mass.  The Ricci tensor is

\begin{equation} \Phi_{00} = \frac{\sigma_v^2}{r^2}, \end{equation}

\noindent for $a_l r < r_c$ and zero for $a_l r>r_c$.  The Weyl tensor is

\begin{equation} \Psi_0 = \left\{ \begin{array}{lcr}
-\frac{2\sigma_v^2 (x-iy)^2}{r^4} & \quad\quad & a_l r/r_c < 1 \\
-\frac{3 r_c \sigma_v^2 (x-iy)^2}{r^5} & \quad\quad & a_l r/r_c > 1
\end{array} \right. . \end{equation}

We take the cut-off radius to be $3.5$~Mpc, which represents a typical size for a massive cluster.  As in the point mass model, we place the lens at $z_l = 0.45$, sources at $z_s = 0.8$, and set the total mass equal to $2.0 \times 10^{15} M_\odot$.

Baltz et al. \cite{baltz} introduced a smoothly truncated model based on the model of Navarro, Frenk, and White (NFW) \cite{NFW}.  The ordinary NFW models, while the best fits to n-body simulations of cluster formation at all scales, suffer from unbounded total mass in the same way as the un-truncated SIS models.  Baltz et al. introduced a smooth truncation scheme which makes the total mass bounded by introducing a new parameter (the tidal radius, $r_t$) inside of which the NFW and tNFW models agree well.  We consider the $n=1$ truncation scheme, for which matter density is modeled as

\begin{equation} \rho(r_p) = \frac{\delta_c \rho_{\mbox{\scriptsize crit}}}
{\left(\frac{r_p}{r_s} \right) \left( 1 + \frac{r_p}{r_s} \right)^2
\left(1 + \left(\frac{r_p}{r_t} \right)^2\right)}.
\label{baltz:rho}\end{equation}

\noindent Here $r_p = a_l r$ is a proper radius, $\rho_{\mbox{\scriptsize crit}}$ is the critical density at the lens redshift, $r_s$ is a scale radius defined as the peak of $r^2\rho(r)$, and $\delta_c$ is a characteristic density contrast.  The characteristic density contrast is related to a concentration parameter, $c$ by

\[ \delta_c = \frac{200}{3}\frac{c^3}{\log{(1+c)}-c/(1+c)}. \]

For our modeling, we take $r_s = 250$~kpc, $c = 7.315$, and set $\tau = 3c$ which ensures good agreement between the NFW model and the tNFW model within the virial radius \cite{baltz}.  As before, we place the lens at $z_l = 0.45$, sources at $z_s = 0.8$, and the total mass for this parameter choice is $2.0 \times 10^{15} M_\odot$.

The three-dimensional gravitational perturbation is

\begin{eqnarray}
\varphi(x_p)  =  \frac{GM_0}{r_s}\frac{\tau^2}{(1+\tau^2)^2} & \Bigg[ &
 \frac{\pi(\tau^2-1)}{2\tau}
-2\ln\tau + \arctan (x_p/\tau) \Big( \frac{1}{\tau}-\tau-2\frac{\tau}{x_p} \Big) \nonumber \\ ~ &~ &
+\ln\left(\frac{1+(x_p/\tau)^2}{(1+x_p)^2}\right)\Big(
\frac{\tau^2-1}{2x_p} -1\Big) \Bigg] \label{baltz:phi1}
\end{eqnarray}

\noindent with $x\equiv a_l r/r_s$. Here  $M_0 \equiv
4 \pi r_s^3 \delta_c\rho_{\mbox{\scriptsize crit}}$ and $\tau = r_t/r_s$. $\Phi_{00}$ is given by

\begin{equation} \Phi_{00} = -\frac{GM_0 a_l^2}{2r_p(r_p+r_s)^2} \frac{r_t^2}
{r_p^2 + r_t^2}. \label{phi00:baltz} \end{equation}

\noindent  The Weyl tensor component is

\begin{equation} \Psi_0 = -\frac{G M_0 a_l^4 r_t^2}{2(r_s^2 + r_t^2)^2} (x-iy)^2
[ F_1(r_p) + F_2(r_p) + F_3(r_p) ] \label{psi0:baltz}\end{equation}

\noindent for

\[ F_1(r_p) = \frac{6 r_s r_t^2}{r_p^4 (r_p^2 + r_t^2)} - \frac{r_t^2 -
r_s^2} { r_p^3 (r_p^2 + r_t^2)} + \frac{4 r_s}{ r_p^2 ( r_p^2 + r_t^2)} \]

\[ F_2(r_p) =  \frac{3(r_t^2-r_s^2)} {r_p^4(r_p+r_s)} + \frac{r_t^2 - r_s^2}{ r_p^3 (r_p+r_s)^2} - \frac{2r_s}{r_p^3(r_p+r_s)} -
\frac{2r_s}{r_p^2(r_p+r_s)^2}, \]

\[ F_3(r_p) = \frac{3 (r_t^2-r_s^2)}{2 r_p^5} \ln \left[\frac{r_s^2 (r_p^2 + r_t^2)} {r_t^2
(r_p+r_s)^2} \right] - \frac{6 r_sr_t}{r_p^5}
\tan^{-1}\left(\frac{r_p}{r_t} \right). \]

\subsection{Conjugate points and curvature tensors}

The Ricci and Weyl tensor components $\Phi_{00}$ and $\Psi_0$ are the source terms in the optical scalar equations and are thus the ultimate source to the geodesic deviation equations.  Understanding the structure of these sources in important to understanding how the numerical integration of the geodesic deviation vectors proceeds.

We have seen that conjugate points can form along the null geodesics, driven by divergences in $\rho$ and $\sigma$ as in Figs.~\ref{rhosig:pm} and \ref{ab:fig}.  Whether the optical scalars have divergences between the observer and source depends on the size of the perturbation along the given geodesic.  In general, the unphysical point mass model has higher peaks in the Weyl tensor perturbation than more physical models with matter density.

\begin{figure}[h]
\begin{center}$
\begin{array}{cc}
\includegraphics[width=2.5in]{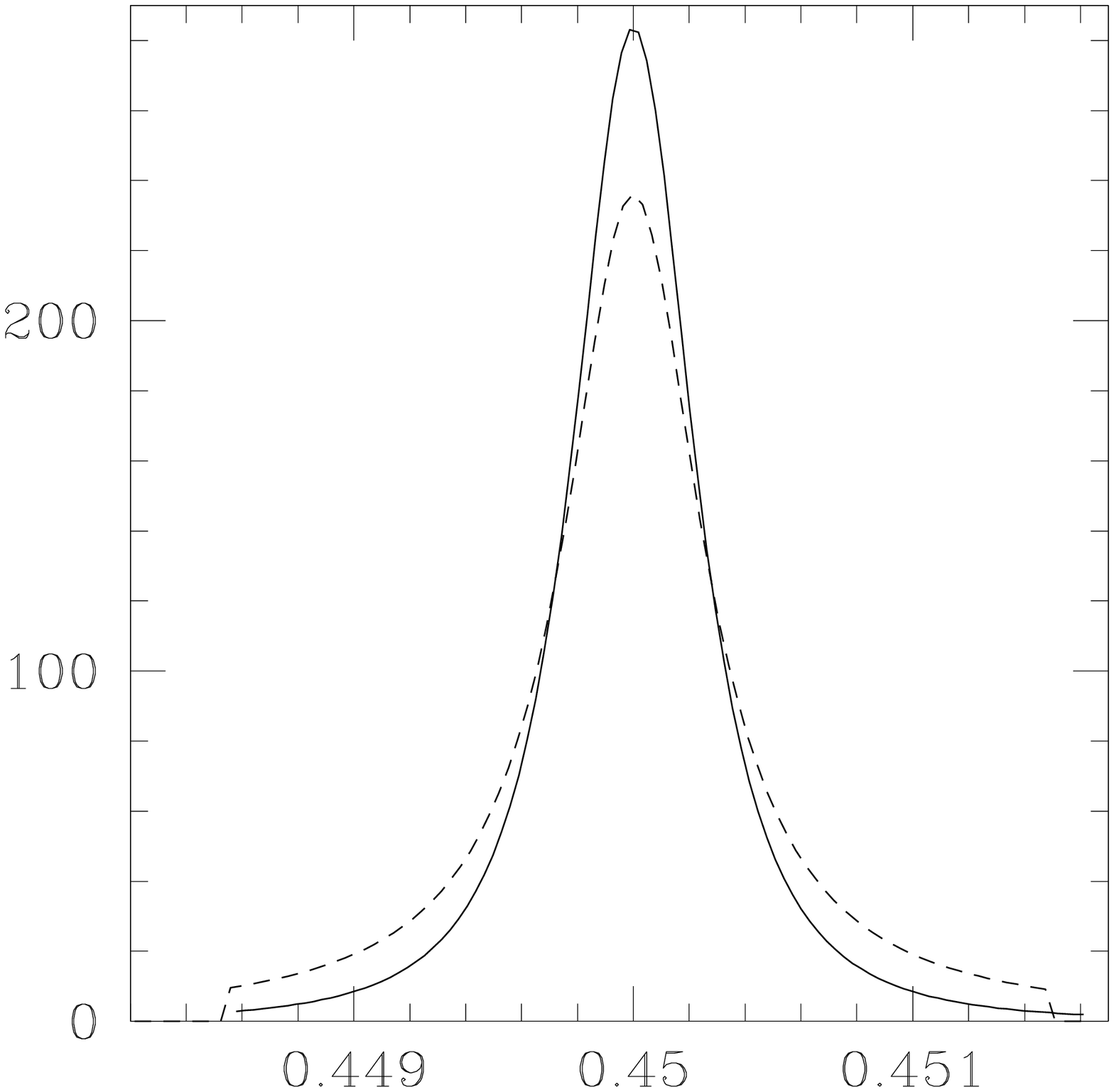} &
\includegraphics[width=2.5in]{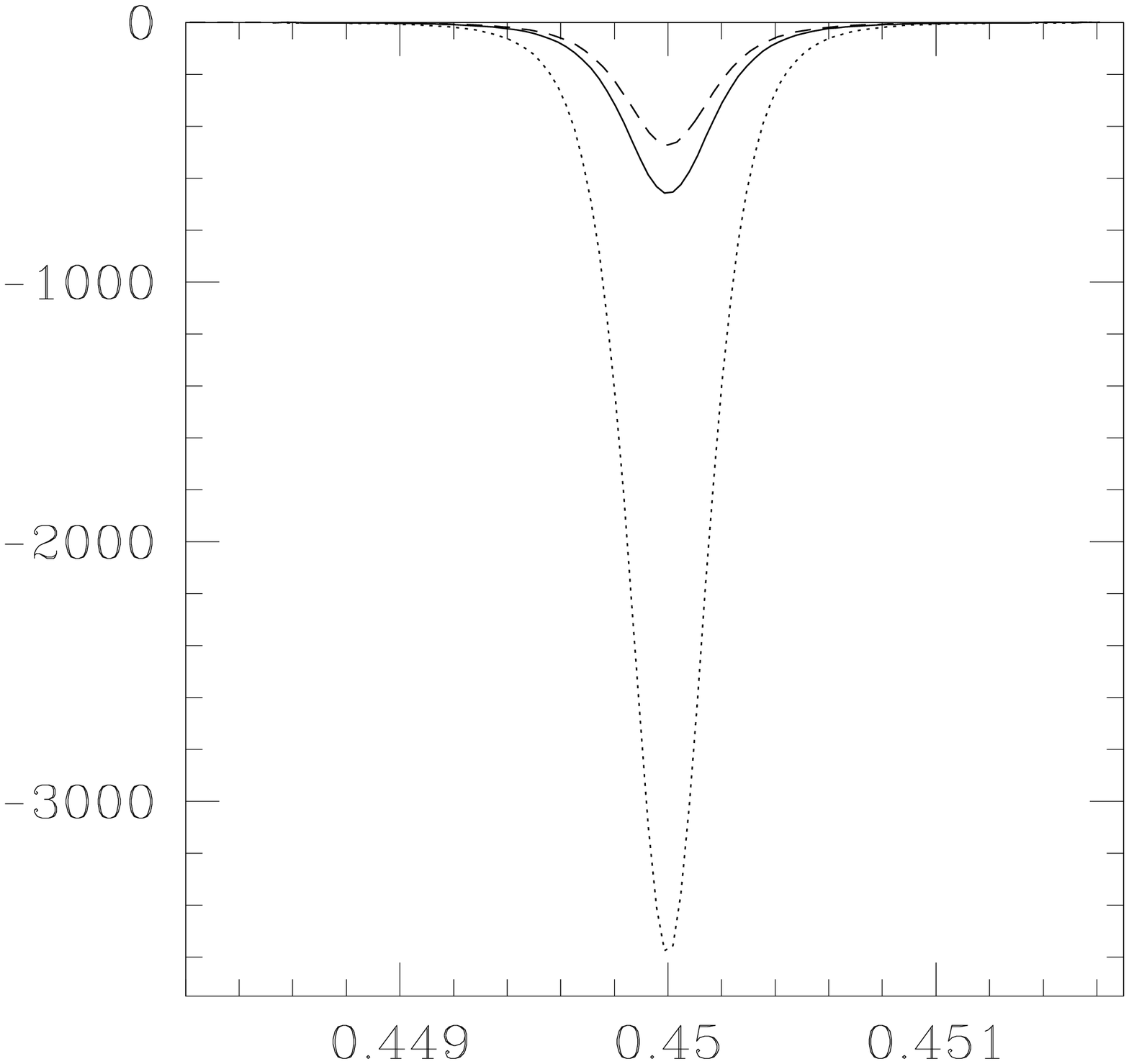}
\end{array}$
\end{center}
\caption{The Ricci ($\Phi_{00}$, left) and Weyl ($\Psi_0$, right) tensor components along a null geodesic in the $\hat x-\hat z$ plane that makes an angle of $120$ arc sec from the $\hat z$ for point mass (dotted), hard-truncated SIS (dashed) and smoothly truncated NFW (solid) models each with a total mass of $2\times10^{15}$ solar masses.  The curvature tensors are scaled into dimensionless units using the age of the universe and are plotted against the redshift.  The hard-truncation of the SIS model shows in the Ricci tensor plot as a step-function-like drop to zero.   \label{ricciandweyl:fig} }
\end{figure}

Figure~\ref{ricciandweyl:fig} shows the Ricci and Weyl tensor terms for the three models as a function of redshift along a null geodesic in the $\hat x$-$\hat z$ plane whose angle of observation is $\theta = 120$ arc seconds.  Both curvature tensors are scaled into dimensionless units. We see that compared with the more physical models considered, that the point mass Weyl tensor is more strongly peaked, leading to more geodesics with conjugate points. The discrete jump in the Ricci tensor term for the hSIS model is due to the matter discontinuity in the hard truncation.

\subsection{Comparisons of thick and thin lens images}

We choose the observer to lie along on the $+\hat z$ axis and the lens to be centered at the origin.  Because of axial symmetry, choosing the source, lens center, and observer to lie in the $\hat x$-$\hat z$ plane makes the Weyl tensor real along the null geodesic connecting the observer and source.  Further, a circular source will appear as an ellipse at the observer. We restrict ourselves to the typical statistical, weak-lensing observations, where sources generally lie outside the critical curve of the lens.  For these sources, their projected image ellipse at the observer will have the semi-major axis aligned with the $\hat y$ direction.

For a given model, we can compute the area and ratio of the semi-major to semi-minor axis for the observed elliptical image using either the thin-lens version of the model and Eqs.~\ref{tlratio} and \ref{tlarea}, or the ratio and areas from integrating the geodesic deviation vectors and using the definitions in Eqs.~\ref{ratio:def} and \ref{area:def}.  We do this by setting an observation angle at the observer, $\theta$, and either applying the thin-lens formulas, or tracing a null geodesic back in time that makes that initial angle with the $\hat z$ axis at the observer until it reaches the source redshift, integrating the optical scalars backwards in time along that path.  The geodesic deviation equations are integrated forwards in time (from source to observer) along the null geodesic using stored values of the optical scalars.

\begin{figure}[h]
\begin{center}
\includegraphics[width=4.5in]{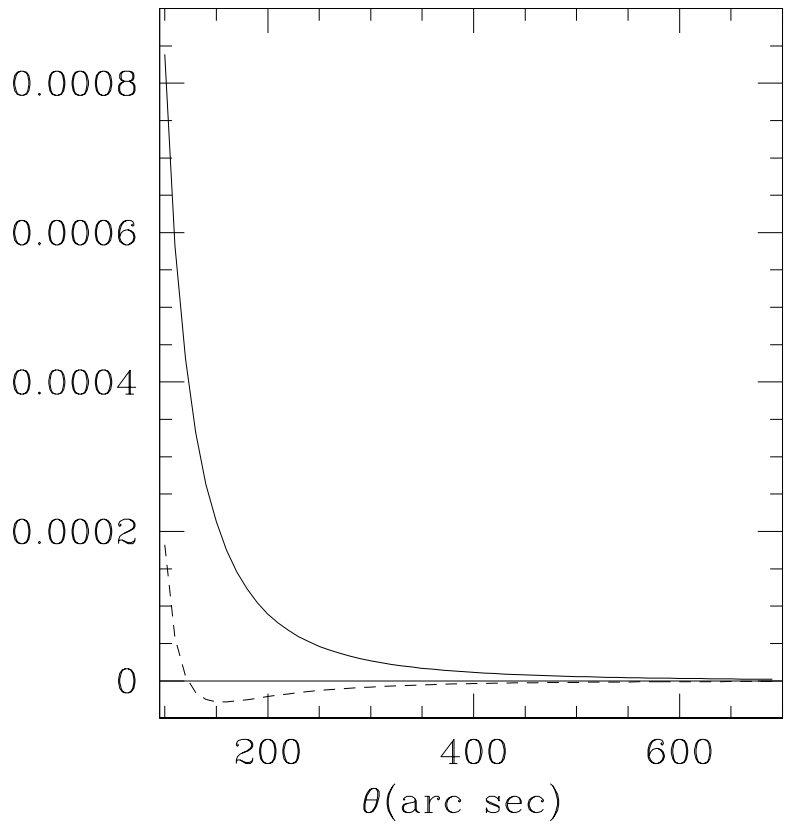}
\end{center}
\caption{The relative error in the ratio of axes (solid) and area (dashed) for a point-mass perturbation where the total mass is $2.0\times 10^{15}$ solar masses.\label{pm:fig} }
\end{figure}

\begin{figure}[h]
\begin{center}
\includegraphics[width=4.5in]{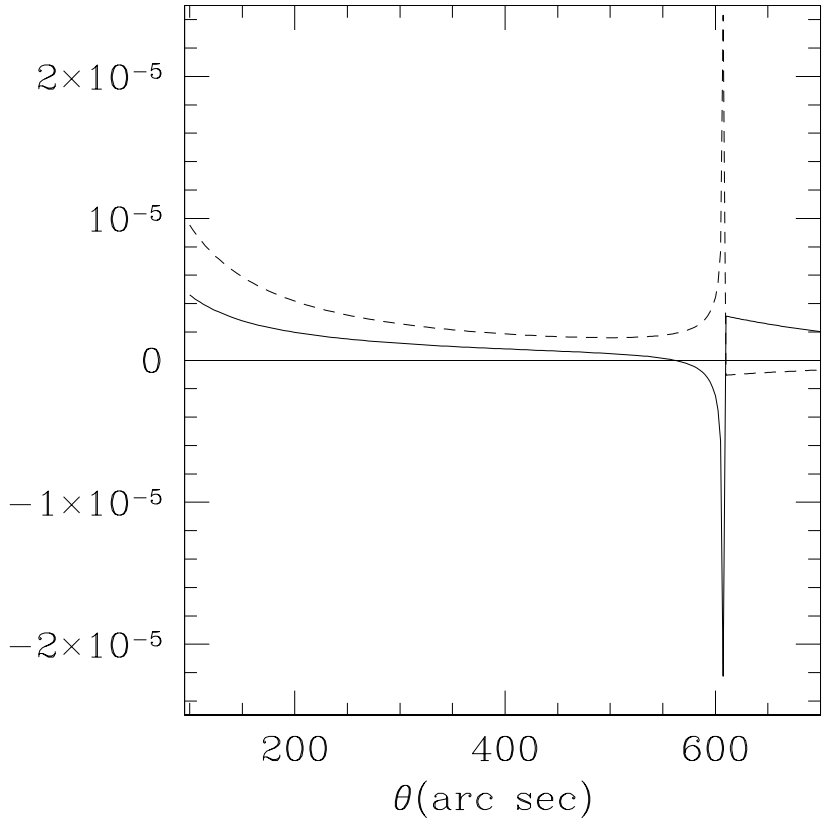}
\end{center}
\caption{The relative error in the ratio of axes (solid) and area (dashed) for a SIS perturbation with hard truncation where the total mass is $2.0\times 10^{15}$ solar masses.  The truncation radius is set at $3.5$~Mpc and the velocity dispersion of the cluster is $1100$~km~s$^{-1}$. The peak in the error profile is not well resolved in this figure and is shown in Fig.~\ref{sis1:fig}.  This peak is located at the truncation radius. \label{sis:fig} }
\end{figure}

\begin{figure}[h]
\begin{center}
\includegraphics[width=4.5in]{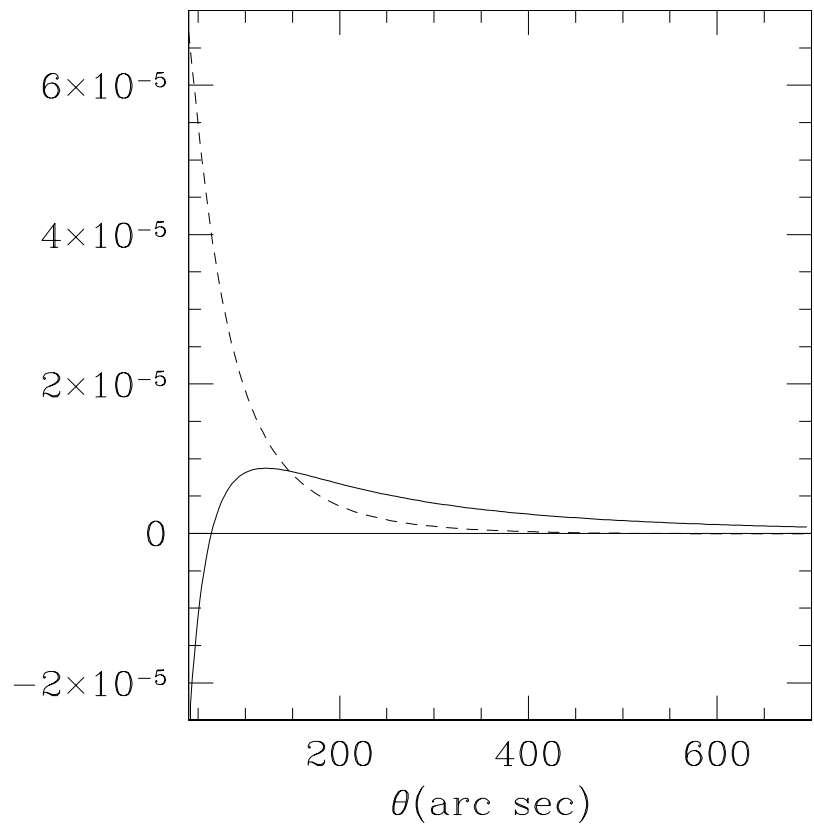}
\end{center}
\caption{The relative error in the ratio of axes (solid) and area (dashed) for a smoothly truncated NFW perturbation where the total mass is $2.0\times 10^{15}$ solar masses.  The scale radius is set at $250$~kpc, the concentration parameter is set at $c=7.3$, and the model's tidal radius is set at $r_t = 3\times c \times r_s$.\label{tnfw:fig} }
\end{figure}

We can then compute a relative difference between the two predictions: $\Delta A / A_p = 1-A_{tl}/A_{p}$ and $\Delta R / R_p = 1-R_{tl}/R_p$.  Taking the thick-lens values as the true values, these two relative differences indicate an error introduced by modeling the system as a thin lens.  Figures \ref{pm:fig}, \ref{sis:fig} and  \ref{tnfw:fig} show the relative errors in the thin-lens predicted areas and axis ratios for the point mass, hSIS, and tNFW models. We see in general that the errors are very small.

In each case, the errors are largest at small radii.  The errors generally grow as a null geodesics approach the Einstein ring radius, where there is a conjugate point.  At this conjugate point along the geodesic, and at the caustic in the lens mapping, both methods break down.

The hSIS model considered here has a matter discontinuity at $3.5$~Mpc where the matter density is (unphysically) set to zero.  This corresponds to about $607.8$~arc sec for our lens and observer positioning.  Inside this radius, where there is matter density, we see general agreement between the hSIS and tNFW models: the thin-lens versions of both models predict smaller axes ratios and areas (positive relative errors) for observation angles away from the Einstein ring angle.  The dip in the ratio of axes in the tNFW model does not appear in the hSIS model.  The difference in the two error profiles is due to the different internal mass structures of the models.

\begin{figure}[h]
\begin{center}
\includegraphics[width=4.5in]{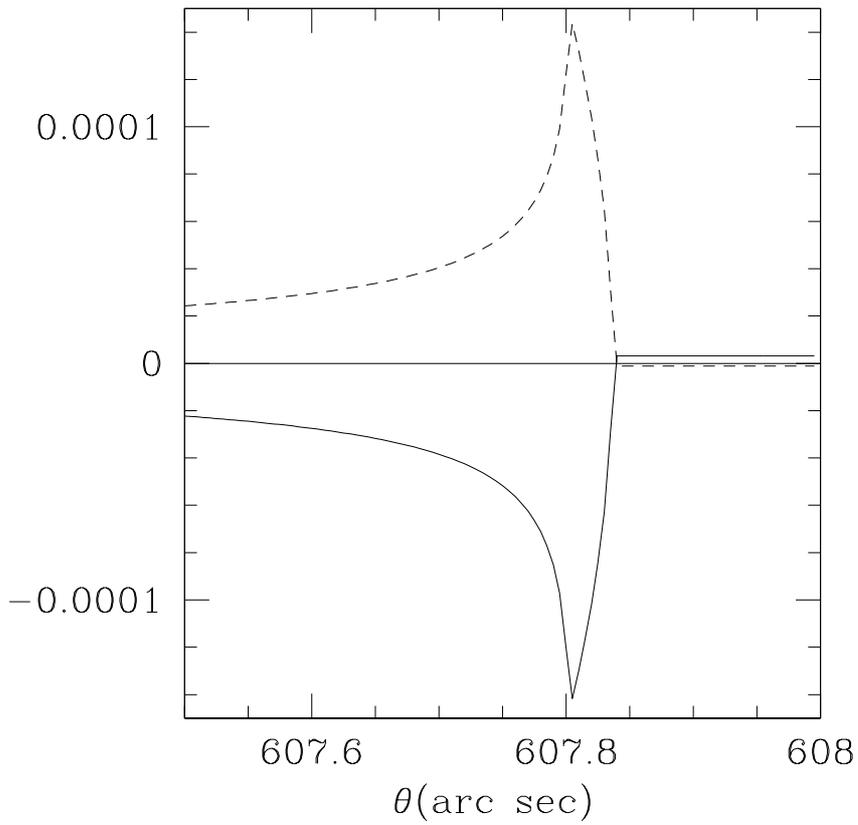}
\end{center}
\caption{The relative error in the ratio of axes (solid) and area (dashed) for the hard truncated SIS perturbation in Fig.~\ref{sis:fig}.  The hard truncation radius of $3.5$~Mpc corresponds to location of the peak in the error profile. Outside this radius, the null geodesics see a point mass perturbation.  This figure shows the consistency in the error plots of the hard truncated SIS and the point mass perturbations. \label{sis1:fig} }
\end{figure}

Past the truncation radius, the hSIS model becomes the point mass model.  Figure~\ref{sis1:fig} shows the errors in the ratio and area for the hSIS model near the truncation radius.  We see that for light rays that do not pass through any of the matter distribution, the errors from Fig.~\ref{sis1:fig} are the same as for the point mass model, Fig.~\ref{pm:fig}: small and positive for the ratio, small and negative for the area.  This result is an internal test of consistency for our computational code.  The peak in the error for the hSIS model represents rays that barely clip the matter distribution. In the case of the smooth, thick-lens approach, these rays remain outside the matter distribution except for a very brief segment.  Since hard truncation is fundamentally unphysical in nature, the peak in the hSIS model is not a physical feature, and it should not influence properly designed lensing studies.  This peak simply indicates an artifact of the unphysical matter modeling.

For the most physical model considered, the smoothly truncated NFW, we see that the thin-lens approximation very slightly underestimates the area and ratio of elliptical images at observation angles away from the Einstein ring angle.

%________________________________________________________

\section{Non-symmetric thick lens} \label{nonsym:sec}

Next, we consider how to reconstruct elliptical images in the thick lensing approach in general.  In the axially symmetric case, one knows, {\emph{a priori}}, the two specific eigen-directions in the cross section of the bundle of light that connects the observer and source.  For a source, observer, and lens center that are co-planar with the $\hat x$-$\hat z$ plane, these are vectors that point in the $+ \hat x$ direction, and $+\hat y$ direction. Note that for a real $\Psi_0$, the shear, $\sigma$, is real, and the geodesic deviation vector equations, Eq.~\ref{geodev1}, uncouple.  If the source lies outside the critical curve in the lens plane, the $+\hat x$ oriented vector will shrink in proportion to the $+\hat y$ oriented vector from the source to observer.  That the geodesic deviation equation uncouples so simply allows one to have a natural, simple, basis set for determining shapes.

In the case of a general lensing configuration, there are still two eigen-directions in the bundle of the light from the source to observer that could form a similar basis pair to the axially symmetric case.  Unfortunately, one does not know these vectors in advance.  The best method is to follow three geodesic deviation vectors from the source to the observer along each null geodesic, and to reconstruct the image ellipse from them.

A general ellipse can be described by

\begin{equation} Y^a(t) = Y^1 (t) e_x^a + Y^2(t) e_y^a,
\label{ellipse}
\end{equation}

\noindent with

\begin{eqnarray} Y^1(t) &=& L_+ \cos(t) \cos(\delta) - L_- \sin(t)
\sin(\delta) \nonumber \\ Y^2(t) &=& L_+ \cos(t) \sin(\delta) + L_-
\sin(t) \cos(\delta) \end{eqnarray}

\noindent where $L_+$ and $L_-$ are the semi-major and semi-minor
axes of the ellipse, $\delta$ is the angle the semi-major axis makes
with $e_x^a$, and $t$ is a parameter that runs from $[0,2\pi)$ \cite{fkn1}.

If $q_i^a = a_i e_x^a + b_i e_y^a$, $q_j^a = a_j e_x^a + b_j
e_y^a$, and $q_k^a = a_k e_x^a + b_k e_y^a$ are three Jacobi fields, denoted collectively by $q_\gamma^a$, for each geodesic deviation vector component, we define

\begin{equation}\hat a_\gamma \equiv \lim_{s\rightarrow 0} \frac{a_\gamma}{s} \quad\quad \& \quad\quad \hat b_\gamma \equiv \lim_{s\rightarrow 0} \frac{b_\gamma}{s},\end{equation}

\noindent to obtain a limiting, projected vector for the image ellipse.  Then we have six equations for six unknowns that determine the observed, projected ellipse:

\begin{eqnarray} \hat a_\gamma &=& (L_+) \cos(t_\gamma) \cos(\delta) - (L_-) \sin(t_\gamma) \sin(\delta), \nonumber \\ \hat b_\gamma &=& (L_+) \cos(t_\gamma) \sin(\delta) + (L_-)\sin(t_\gamma) \cos (\delta),  \label{shape} \end{eqnarray}

\noindent where $t_\gamma = (t_i, t_j, t_k)$ indicate the angle each of the three vectors
$q_\gamma^a = (q_i^a, q_j^a, q_k^a)$ make along the ellipse measured from the semi-major axis.  Therefore, by integrating the geodesic deviation equations, Eqs.~\ref{geodev1}, simultaneously for three Jacobi fields, one can determine the lengths of the semi-major and semi-minor axes as well as the orientation of the ellipse from Eqs.~\ref{shape} for general asymmetric lenses.

As a first test this approach, we model a non-symmetric thick lens by superimposing two tNFW profiles whose centers are offset and whose total combined mass is $2\times 10^{15}~M_\odot$. We place the observer on the $\hat z$ axis as before and a series of circular sources in a source plane at constant $-\hat z$ such that the sources lie at a redshift of $z=0.8$.

The center of the first and slightly more massive cluster, $\rho_{m1}$, is positioned on the $\hat z$ axis at a redshift of $z=0.448$.  The center of the second clump, $\rho_{m2}$, lies in the $\hat x - \hat z$ plane $20$~arc sec from the $\hat z$ axis in the direction of $+\hat x$.  This second clump is at a redshift of $z=0.452$.  With this positioning, the tidal radii of the two clusters barely overlap -- the center to center proper distance (at a redshift of $0.45$) is approximately $1.1$ times the sum of the tidal radii $r_{t1} + r_{t2}$ (also proper distances). Table \ref{2lens:table} indicates the parameters used for the model.

\begin{figure}[h]
\begin{center}
\includegraphics[width=4.5in]{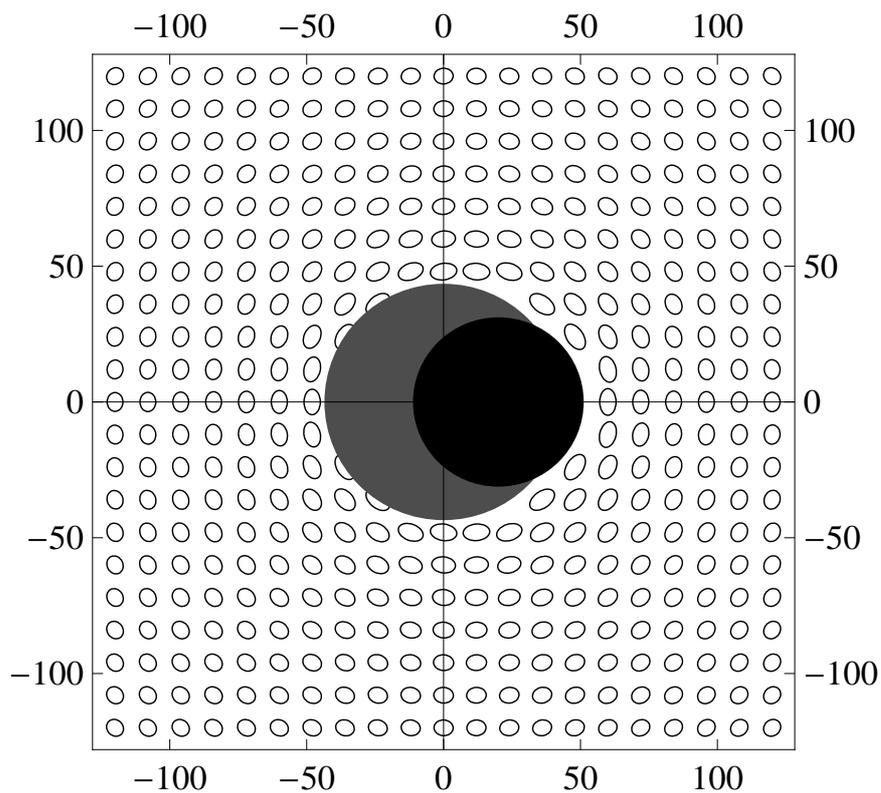}
\end{center}
\caption{The images of a sky of circular sources lensed by two thick lenses recovered by the thick-lens approach.  The two clusters are tNFW models at redshifts of $0.448$ and $0.452$.  The closer lens is placed on the optical axis and has a total mass of $1.4 \times 10^{15}$ solar masses, while the more distant cluster is placed $20$ arc sec towards $+\hat x$ and has a total mass of $0.6 \times 10^{15}$ solar masses.  The gray and black disks represent the projection of the two clusters scale diameters $2\times r_s$ on the sky.  Each cluster has a tidal diameter extending past $600$ arc sec. \label{2lenssky:fig} }
\end{figure}

Figure \ref{2lenssky:fig} shows what would be the elliptically oriented images of a background sky of circular sources.  The size of the ellipses drawn here is magnified to show the effects of the lensing. Since we only consider objects whose line of sight is greater than $40$ arc sec away from a position $+10$ arc sec rightward of the origin, it is reasonable that the images are roughly elliptical and not arcs. Also shown are disks corresponding to the scale radii of both distributions. Disks for the tidal radii would extend past the borders of the figure.  One sees the recovery of elliptical shapes, demonstrating the ability to use the thick-lens approach to obtain images of the weak lensing of background galaxies.

%_________________________________________________________________

\section{One lens plane analysis of thick lenses}

Most wide-field surveys used in gravitational lensing detect lensing clusters in portions of the sky with no spectroscopic observations, and photometric redshift approximations are utilized.  Using photometric redshifts only, in the case considered above, where the two lensing clusters overlap only at their outmost edges, but are only separated by $0.004$ in redshift, one could not tell that there are two clusters separated along the line of sight.  Instead, one would model this situation with one lens plane.

Our thick lens approach to weak lensing allows us to examine the error introduced by modeling two separate clusters as being in one lens plane.  For the thin lens approximation, we assume that the projected lensing potential is the sum of two projected tNFW model potentials, $\psi = \psi_1 + \psi_2$, where both lensing potentials are at a distance corresponding to a redshift of $0.45$.  We use the formulas for the tNFW lensing potential the appendix of Baltz et al. \cite{baltz}.  We then compute the Jacobian matrix of the lens map, and use the eigenvalues and eigenvectors of the thin-lens mapping inverse Jacobian matrix to find the shape of the elliptical source following standard references.

Against the thin lens in a single lens plane, we take the two three-dimensional clusters and separate them along the line of sight and integrate the thick-lens shear and convergence and Jacobi vectors to find the shape of the image.  The observer and center of the more massive cluster are always on the $+\hat z$ axis, and the center of the second, less massive, cluster is always placed in the $\hat x$-$\hat z$ plane at an angular separation of $20$ arc seconds towards $+\hat x$  from the $+\hat z$ axis.  The second cluster is always placed at negative $z$ values, such that the center of the two clusters (along the $\hat z$ axis) is at a redshift of $0.45$.  The redshifts of the two clusters are set according to $z_{1,2} = 0.45 \pm i \times 0.0004$, and the greatest separation of redshift values we consider places the two clusters at redshifts of $0.44$ and $0.46$.  Photometric redshifts certainly could tell at this greatest separation in redshift that there are two separate clusters.  In the modeling for this paper, we use the same $(c, r_s)$ pairing as in Table~\ref{2lens:table} for the two clusters.

\begin{figure}[h]
\begin{center}
\includegraphics[width=4.5in]{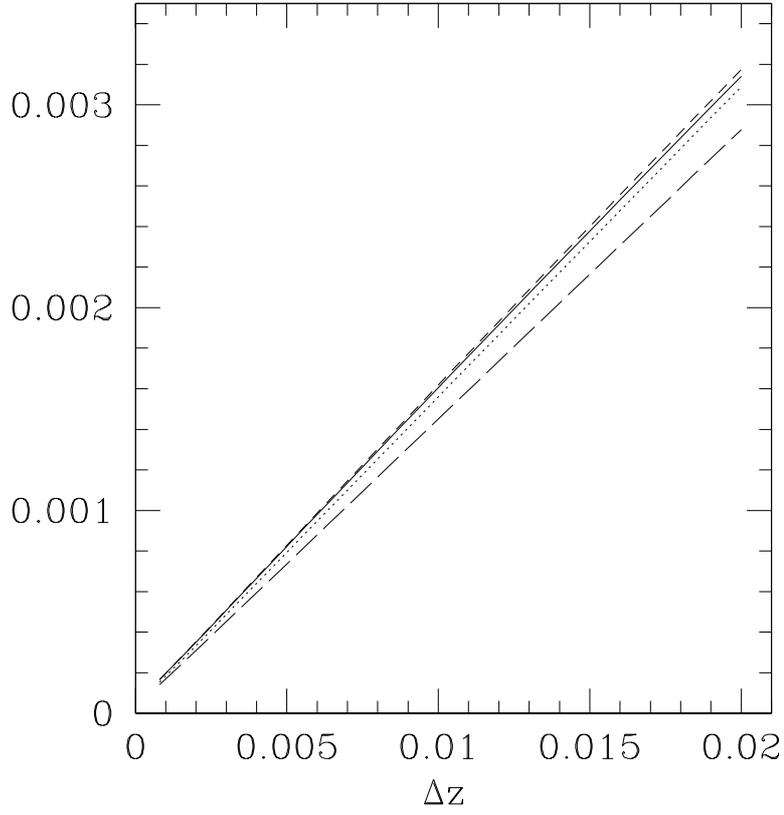}
\end{center}
\caption{The relative error in the projected area as seen by an observer between thick and thin-lens analysis for two thick lenses separated in redshift space by $\Delta z$.  The thin-lens analysis assumes both lenses are at the same redshift, $z=0.45$, which is the center of the redshift separation.  The two lenses have the same tNFW parameters as in Table~\ref{2lens:table}.  The four error curves correspond to four different sky image locations:  $(75,0)$ - solid curve, $(50,50)$ - short dashed curve, $(-50,50)$ - dotted curve, $(-75,0)$ - long dashed curve. \label{2lensarea:fig}}
\end{figure}

\begin{figure}[h]
\begin{center}
\includegraphics[width=4.5in]{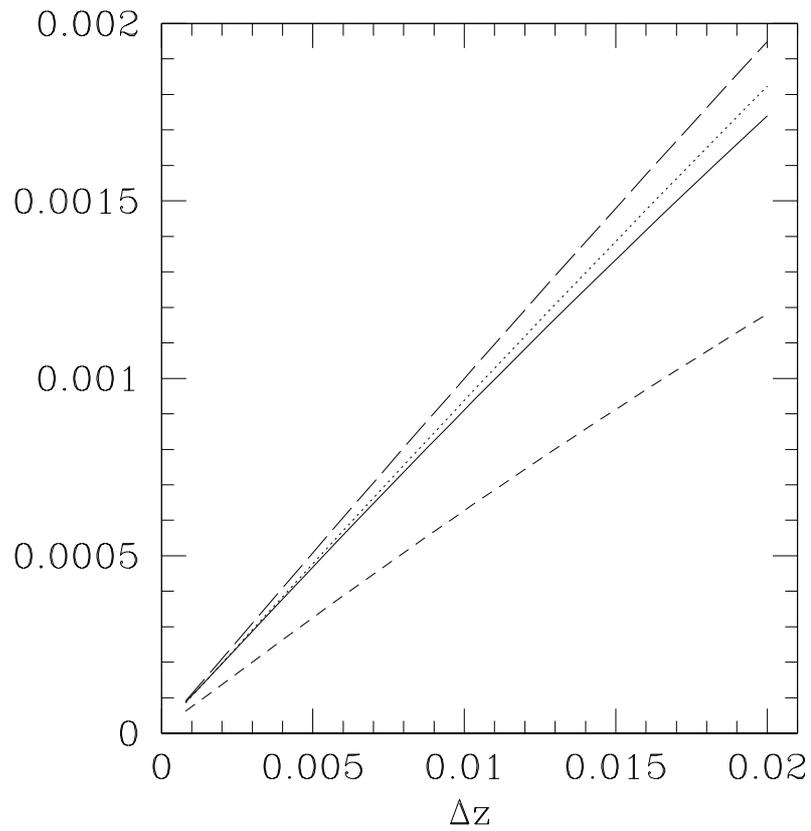}
\end{center}
\caption{The relative error in the projected axes ratio as seen by an observer between thick and thin-lens analysis for two thick lenses separated in redshift space by $\Delta z$, following the same set-up as Fig.~\ref{2lensarea:fig}. The four error curves correspond to four different sky image locations:  $(75,0)$ - solid curve, $(50,50)$ - short dashed curve, $(-50,50)$ - dotted curve, $(-75,0)$ - long dashed curve. \label{2lensratio:fig}}
\end{figure}

This allows us to plot the relative errors in ellipse ratio and area for the thin-lens compared with the thick lens as a function of redshift separation, $\Delta z = z_2-z_1$.  Figure~\ref{2lensarea:fig} shows the relative error in the image area, $1-A_{tl}/ A_p$, plotted against redshift separation for four image locations at $(x,y)$ arc sec separation from origin: $(75,0)$, $(50,50)$, $(-50, 50)$, $(-75, 0)$.  Figure~\ref{2lensratio:fig} shows the relative error in image axes ratio, $1-R_{tl}/R_p$, against redshift separation for the same image locations on the sky.  In both cases, three geodesic deviation vectors were used to first determine the projected ellipse seen by the observer, and then the ratio and projected area of the ellipse was measured.

\begin{figure}[h]
\begin{center}$
\begin{array}{cc}
\includegraphics[width=2.5in]{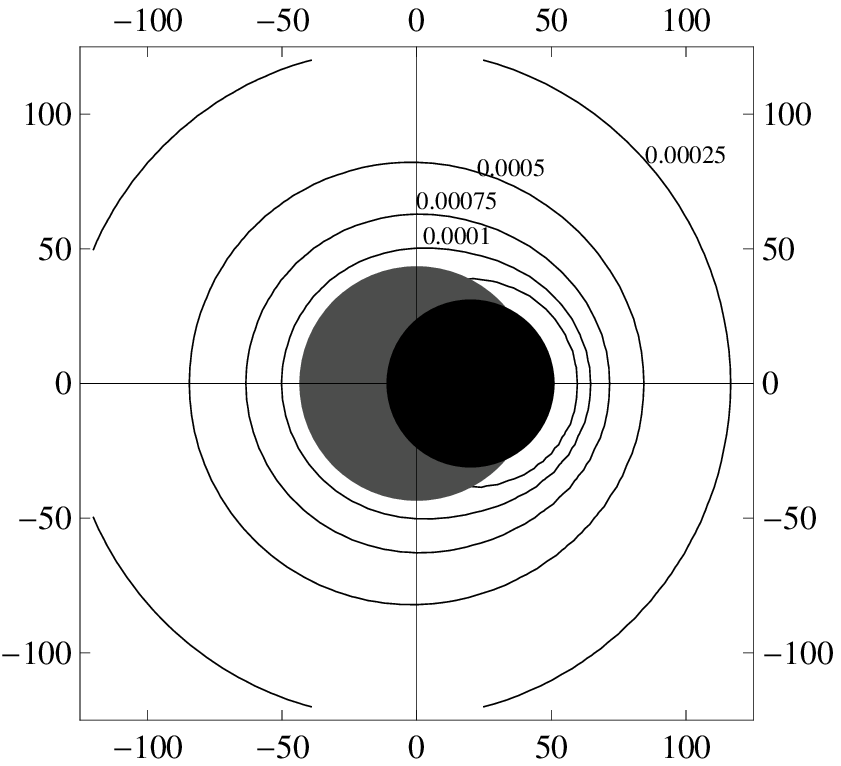} &
\includegraphics[width=2.5in]{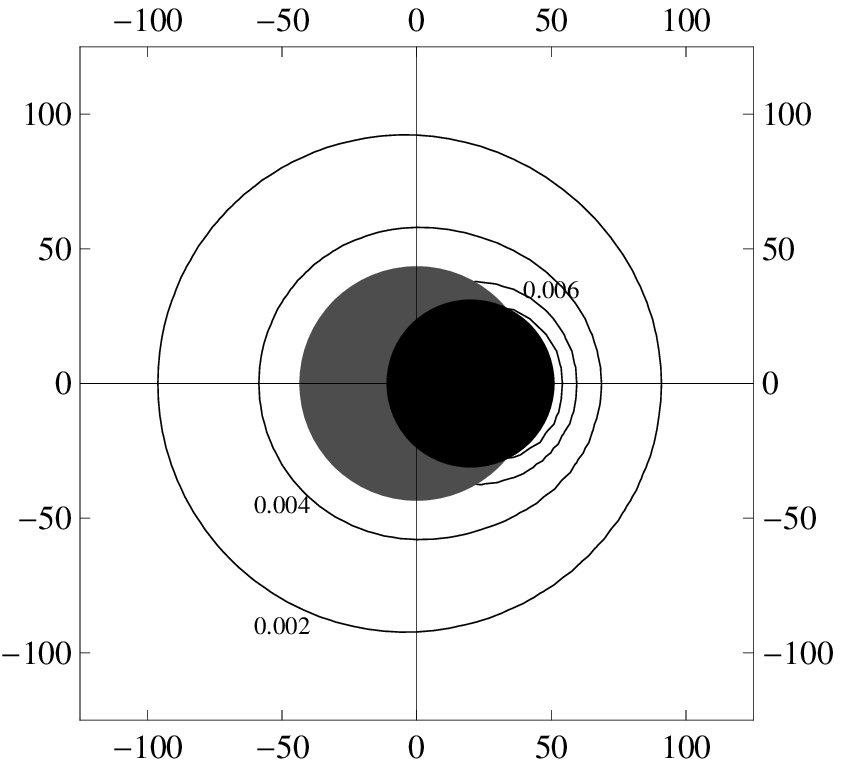}
\end{array}$
\end{center}
\caption{A contour plot of the error in the projected area between the thin-lens and thick-lens modeling of two lensing clusters.  In the left panel, the clusters are positioned redshifts of $z=0.448$ and $z=0.452$.  In the right panel, the redshifts are $z=0.44$ and $z=0.46$.  The disks indicate the projection of the size of the scale diameter $2\times r_s$.\label{contour:area} }
\end{figure}

\begin{figure}[h]
\begin{center}$
\begin{array}{cc}
\includegraphics[width=2.5in]{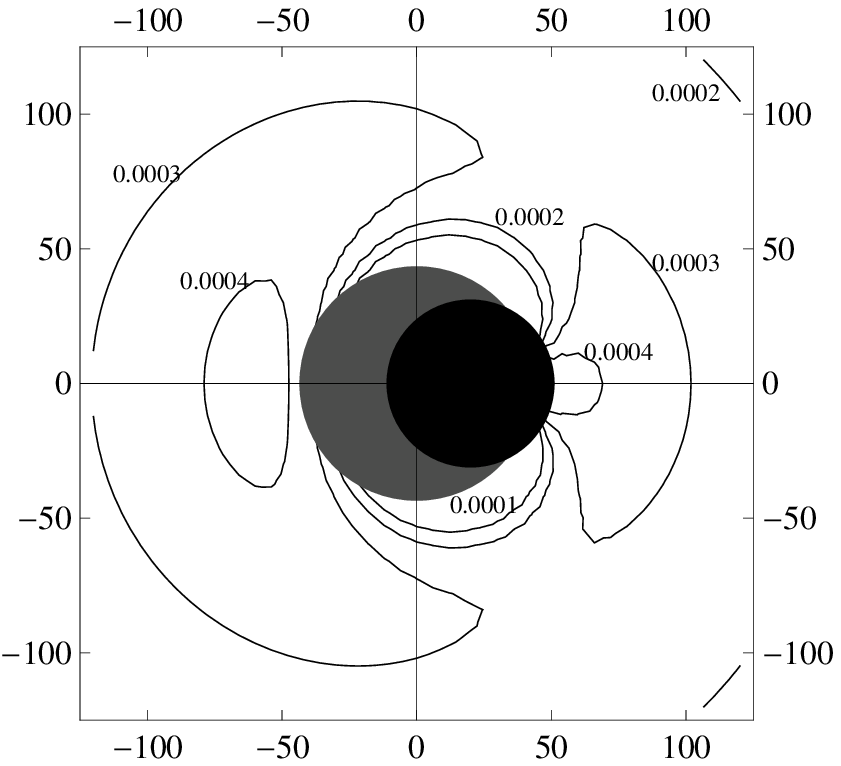} &
\includegraphics[width=2.5in]{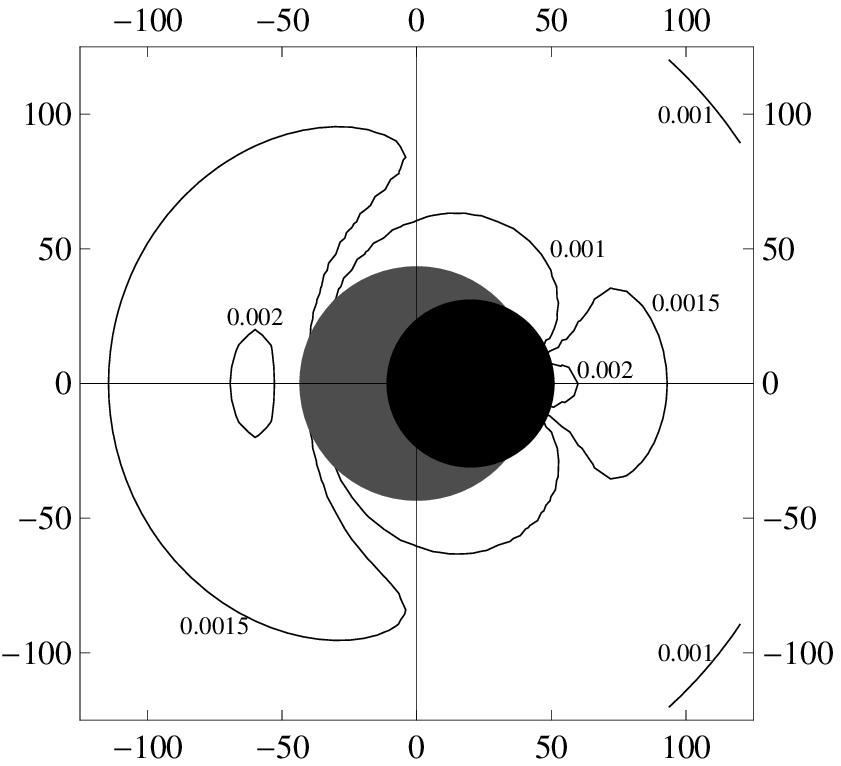}
\end{array}$
\end{center}
\caption{A contour plot of the error in the projected axes ratio between the thin-lens and thick-lens modeling of two lensing clusters.  In the left panel, the clusters are positioned redshifts of $z=0.448$ and $z=0.452$.  In the right panel, the redshifts are $z=0.44$ and $z=0.46$.  The disks indicate the projection of the size of the scale diameter $2\times r_s$. \label{contour:ratio} }
\end{figure}

Figures~\ref{contour:area} and \ref{contour:ratio} show contour plots of the error in the image ellipse's area and axes ratio when the two lenses are placed close together, at redshifts of $0.448$ and $0.452$, or far apart, at redshifts of $0.44$ and $0.46$.

We see that the value of the relative error in area and axes ratio is positive.  In all cases, the value of the error at a given point increases as the two clusters are separated.  However, even at very large separation in redshift, the overall relative error in the axes ratio and image area remains small -- less than $1\%$ for the area and less than $0.5\%$ for the ratio.

%____________________________________________________________

\section{Discussion}

Previous papers on gravitational image distortion from the non-perturbative stand-point did succeed in laying the ground-work and theory for weak gravitational lensing \cite{perlick}, \cite{fkn1}, \cite{fkn2},  \cite{fo}. However, each of these papers examined non-physical lenses, did not integrate along null geodesics, and used somewhat opaque language, familiar to the relativity community, but unfamiliar to the practicing astrophysics community.  This paper extends those findings by working through the details of physical lenses, with both Ricci and Weyl curvature, and states results in terms of typical, ground-based, statistical weak gravitational lensing studies.

We provide a practical program for integrating the gravitational weak lensing image distortion for thick lenses.  We draw together the elements of the Newman-Penrose spin-coefficient formalism, known to a subset of relativists, and present them in a simplified form applicable to weak metric perturbations of a background metric which is conformal to the flat Robertson-Walker metric. This program is

\begin{enumerate}
\item place the observer, lens, and sources using measured redshifts, the functional form of the scale factor, Eq.~\ref{a1}, and background null geodesics,
\item simultaneously integrate the null geodesic equations, Eqs.~\ref{eqns:motion} and the Sachs equations for the optical scalars, Eqs.~\ref{sachs:rho} and \ref{sachs:sigma}, from the observer back in time to the source,
\item integrate a combination of geodesic deviation vectors forwards in time from the source to the observer using Eq.~\ref{geodev1}, and determine the projected shape of the image by dividing the components by the differential parameter, $s$ and taking the limit as $s\rightarrow 0$.
\end{enumerate}

\noindent We demonstrate this method in a variety of scenarios, including different metric perturbations and non-axially symmetric thick lenses.

The traditional weak lensing analysis of a non-axially symmetric lens involves finding the eigenvectors and eigenvalues of the inverse of the Jacobian of the thin-lens mapping. Modern approaches involve solving partial differential equations for the projected gravitational potential \cite{hossein} or the projected mass density \cite{seitz}.  Integrating the true null geodesics has typically been avoided because it was seen as too computationally intensive.

The program for finding image distortion from thick lenses that we outline and implement numerically in this paper demonstrates that thick-lens analysis has become more reasonable.  In fact, the time and computationally intensive tasks of gravitational lensing analysis are image processing, stacking, and shape measurement.  While thick-lens analysis involving integrating null geodesics and geodesic deviation vectors may take longer than the corresponding thin-lens analysis, it is at least an order of magnitude smaller task in computation and time than image processing and shape measurement.

Nevertheless, in the case of lensing systems that are axially symmetric, we find that the thin-lens approximation generates image ellipses whose area and axes ratio is nearly identical to the corresponding thick-lens analysis.  The errors for axially symmetric systems found in this paper are small compared to observational uncertainties.

In practice, one observes the averaged ellipticity of sources in a region of the sky since no individual background galaxy is known to be circular.  For an object with axes ratio $R$, the ellipticity is defined as

\[ |\epsilon| = \frac{1-R}{1+R}. \]

\noindent Averaging source ellipticities carries with it an intrinsic error determined by the number of objects averaged over and the intrinsic ellipticity dispersion, $\sigma_\epsilon$, as $\sigma = \sigma_\epsilon/\sqrt{N}$.  Typical values of $\sigma_\epsilon$ are about $0.3$.  The density of observed images is typically about $20$ useable background objects per square arc minute, as in the DLS survey \cite{wittman}, or higher by up to a factor of 2 in other current ground-based studies.

In analyzing axially symmetric lensing configurations, one typically introduces annular bins for averaging ellipticity measurements.  For instance, one might introduce fifteen annular bins of equal width with the center of the first bin at $100$ arc sec and the center of the last bin at $600$ arc sec, and then one would average the ellipticity of every object in the bin.  Within each bin, there would be an expected number of background objects based on the number of objects per square arc minute and the size of the annular bin.  For a bins at $100$ arc sec and $250$ arc sec, the observational accuracies would be approximately $\sigma \approx 0.027$ and $\sigma \approx 0.017$, respectively, for the DLS survey.

We see that the observational uncertainties are several orders of magnitude larger than the error introduced by using a thin-lens approximation.  Therefore, the thick lens analysis is only of practical benefit if one is attempting to model structure with parameters along the line of sight.  For example, one might attempt to model a galaxy cluster as a triaxial ellipsoid rather than a projection into a two dimensional elliptical matter distribution.

We note that non-truncated SIS models are commonly used in weak gravitational lensing studies based on the thin-lens approximation.  However, truncation is important at the scales of wide-field gravitational lensing analysis -- even large clusters are not expected to contain a significant mass density past two or three Mpc in radius, which is within the field of view for almost all ground-based weak lensing observations of clusters.  In this paper, we show that the error in the thin-lens predicted ellipse area and axes ratio spikes at the truncation radius. This error spike is a needless artifact of poor matter modeling.

Of significance to the lensing community, is the finding in this paper that a single lens plane model of two clusters, separated by a significant distance along the line of sight, produces highly accurate image area and axes ratios.  Typical weak-lensing surveys are not supported by spectroscopic observations of cluster members, but rely instead on photometric redshifts to determine cluster membership.  We show in this paper that the weak lensing image distortion of two clusters that are significantly separated in distance along the line of sight but close together on the sky can be modeled by a single lens plane to high accuracy.

This is a problem in two regards.  First, studies that attempt to determine the number distribution of clusters of different masses at different redshifts that are based on gravitational lensing may under-count the total number of clusters, and over-state their mass.  Second, in studies of structure formation within clusters, if clusters are selected on the basis of weak-gravitational lensing image distortion signals, two well-separated clusters might be mistaken for a single merging object.

These findings argue strongly for spectroscopic follow-up to clusters selected from weak-gravitational lensing surveys to make sure that the cluster members lie at one redshift.

%__________________________________________________________

\section{Conclusions}

We have integrated the optical scalar and geodesic deviation equations along the null geodesics of a perturbed space-time representing real clusters in a Robertson-Walker background.  Our calculations demonstrate the feasibility of analyzing thick gravitational lenses in a non-perturbative scheme and allow an examination of the underlying accuracy of the common-thin lens approximation.

Our principle finding in this context is that the thin-lens approximation is in a sense too good.  Namely, we find that if there are two lenses significantly separated along the line of sight, a single lens plane model of these two lenses reproduces the shapes and orientations of image ellipses with an error less than what one could expect from averaging over the images of close-by uncorrelated background sources.  This implies that spectroscopic observations are important supplements to weak lensing observations.

Our initial examinations of the thin-lens approximation for weak-lensing image distortion shows that a poor truncation scheme like hard truncation can introduce an error that is as large as the inherent thin-lens approximation error in central regions of a properly truncated model.  Since wide-field, ground-based studies of weak lensing do examine the region where the truncation is expected to take place, more careful examinations of the influence truncation on the relative error in area and axes ratio is warranted.

%__________________________________________________________

\section {Appendix: Numerical methods}

In this appendix, we describe details of the numerical calculations in the paper.

It is most convenient (for numerical integration) to scale all the variables by the age of the universe, which we determine by setting $a=1$ in Eq.~\ref{a1} and solving for $\tilde t$.  This has the effect of making distances from the observer to the source have roughly unit size, and the parameter $s$ tends to run from $0$ to about $0.75$ depending on arrangements of source and observer.
The effect of the lens on the path of the null geodesic and also the effect on the optical scalar equations take place over a very short range in $s$, so adaptive step-size methods with error corrections need to be used.  In this paper, we follow a personal adaptation of the Cash-Karp implementation of the Runge-Kutta-Fehlberg 4-5 adaptive step-size method \cite{nrc}.

The null geodesic path and optical scalars, $\rho$ (or preferably $u=1/\rho$) and $\sigma$, are first integrated back in time from the observer to the source plane.  An initial observation angle is set at the observer, and the numerical integration continues until the source plane is reached.  We refine the step size to stop at the source plane $z$ coordinate value with accuracy to one part in $10^{12}$.  As the path and optical scalars are integrated, the values of the optical scalars are passed into an array $(s, \rho, \sigma)$ at every step along the numerical integration.

Since $u = 1/\rho$ is well defined at the observer, an adaptive step-size algorithm would naturally take large steps near the observer.  However, $\rho$ divergences at the observer, and this divergence is responsible for making the geodesic deviation vectors vanish at the observer. Therefore, we force the adaptive step-size algorithm to take many small steps near $s=0$ in the integration of the path and optical scalars away from the observer towards the source.  For most rays, we store approximately 25,000 to 45,000 steps in the $(s, \rho, \sigma)$ array, with the vast majority of these steps either near the observer or near the lens.

We then integrate the geodesic deviation equations back from the source to the observer, setting the initial conditions at the source to define a circular cross-section of the pencil.  For this integration, we again use an Runge-Kutta-Fehlberg 4-5 adaptive step-size algorithm.  We interpolate values for $\sigma(s)$ and $\rho(s)$ from the stored arrays by fitting a fourth order polynomial to the nearest four stored data values.  Both optical scalars are smooth functions along the integration, although $\rho$ diverges in the limit $s \rightarrow 0$.

Working to double precision throughout, the total accumulated error in the numerical integration of null geodesic and optical scalars (from the observer to source) is less than one part in $10^{14}$.  The total accumulated error in the numerical integration of the geodesic deviation vector components is estimated from the adaptive step-size algorithm as less than one part in $10^{12}$, and the geodesic deviation vector component integration typically proceeds in less than 3000 steps.  However, further numerical analysis of our interpolation indicates that the interpolation might add an error of up to one part in $10^{10}$.

If $s_l$ is the value of the parameter of $s$ along the geodesic at the lens, we find that there is essentially no change in the projected components of the geodesic deviation vectors, $a/s$ and $b/s$, for $s < s_l/20$.  Therefore, we take the projected limits by setting a floor value of $s = s_{tiny}$ close to $s=0$ and stopping the numerical integration there:

\[ \lim_{s\rightarrow 0} \frac{a}{s} \approx \frac{a(s_{tiny})}{s_{tiny}}. \]

\noindent We note that the choice of $s_{tiny}$ is somewhat arbitrary, but that the best choice is related to the maximum step-size set in integrating $u$ near the observer, as well as the value of $s$ where the forced small step size ends.

For the non-axially symmetric cases, we integrate the null geodesic equations and optical scalar equations as before, but then simultaneously integrate three geodesic deviation vectors back from the source to the observer.  The initial conditions for the triplet of geodesic deviation vectors are taken such that if the lens was not present the triplet of geodesic deviation vectors would be $(1,0)$, $(-1/2, \sqrt{3}/2)$, and $(1/3, 2\sqrt{2}/3)$ at the observer.  This choice of three unrelated vectors insures that Eqs.~\ref{shape} are not linearly dependent.  Solving Eqs.~\ref{shape} is somewhat difficult numerically because of the trigonometric functions.  Modifying a standard Newton-Raphson method to insure that the angular variables do not make large changes (passing $2\pi$, for example) is critical in obtaining a solution.  Both Maple and Mathematica can find solutions quickly, especially if the initial guess is adequate, but both give angular values for $t_i$ and $\delta_i$ in unusual multiples of $\pi$.

Throughout this paper, error calculations and figures are produced from C++ code based on appropriate algorithms from Numerical Recipes \cite{nrc}.  All calculations were independently coded in Mathematica for internal consistency checks.

%___________________________________________________________

\begin{acknowledgements}
LB thanks the Bridgewater State College Adrian Tinsley Program for
Undergraduate Research for a Summer Grant that enabled his
participation in this project.  Work by TPK was supported by a Bridgewater State University Faculty and Librarian Research Grant.
\end{acknowledgements}

%_________________________________________________________

% FIGURES and TABLES

\begin{table} \begin{center}\begin{tabular}{cccccc}
\hline z &  $\theta$ (arc sec) & $c$ & $r_s$ (kpc) & $M_{tot}$ ($10^{15}~M_\odot$) & $\Delta$ \\
\hline ~&~&~&~&~\\
0.448 & 0.0 & 6.43 & 250 & 1.4 & --\\
0.452 & 20.0 & 6.76 & 180 & 0.6 & 1.1\\
\hline
\end{tabular} \caption{Parameters for two superimposed truncated NFW profile matter distributions including redshift and angle from the $\hat z$ axis of the center of the distribution used for Fig.~\ref{2lenssky:fig}. The model parameters and total mass of the cluster (in units of $10^{15}$ solar masses) are also given.  $\Delta \equiv \Delta r_p / (r_{t1} + r_{t2})$ indicates the ratio of the proper distance between the centers of the two distributions and the sum of tidal radii.\label{2lens:table} }
\end{center}
\end{table}

%____________________________________________________________

%______________________________________________________________

%_________________________________________________________________________

%_________________________________________________________________________

%_________________________________________________________________________

%_________________________________________________________________________

%_________________________________________________________________________

%_________________________________________________________________________

%_________________________________________________________________________

%_________________________________________________________________________

%_________________________________________________________________________

%_________________________________________________________________________

%_____________________________________________________________________

\end{document}